\documentclass[twocolumn]{aastex631}

\usepackage{CJKutf8}
\usepackage{amsmath}

\begin{document}

\title{AGN jet-inflated bubbles as possible origin of odd radio circles}

\correspondingauthor{Yen-Hsing Lin}
\email{julius52700@gmail.com}

\author[0000-0002-6082-3813]{Yen-Hsing Lin}
\affiliation{Institute of Astronomy, National Tsing Hua University, Hsinchu 30013, Taiwan}

\author[0000-0002-1311-4942]{H.-Y. Karen Yang}
\affiliation{Institute of Astronomy, National Tsing Hua University, Hsinchu 30013, Taiwan}
\affiliation{Physics Division, National Center for Theoretical Sciences, Taipei 106017, Taiwan}
\email{hyang@phys.nthu.edu.tw}





\begin{abstract}

Odd radio circles (ORCs) are newly discovered extragalactic radio objects with unknown origin. In this work, we carry out three-dimensional cosmic-ray (CR) magnetohydrodynamic simulations using the FLASH code and predict the radio morphology of end-on active galactic nucleus (AGN) jet-inflated bubbles considering hadronic emission. We consider CR proton (CRp)-dominated jets as they tend to inflate oblate bubbles, promising to reproduce the large inferred sizes of the ORCs when viewed end-on.
We find that powerful and long-duration CRp-dominated jets can create bubbles with similar sizes ($\sim 300-600$ kpc) and radio morphology (circular and edge-brightened) to the observed ORCs in low-mass ($M_{\rm vir}\sim 8\times 10^{12} - 8\times 10^{13}~M_\odot$) halos. Given the same amount of input jet energy, longer-duration (thus lower-power) jets tend to create larger bubbles since high-power jets generate strong shocks that carry away a significant portion of the jet energy. The edge-brightened feature of the observed ORCs is naturally reproduced due to efficient hadronic collisions at the interface between the bubbles and the ambient medium.
We further discuss the radio luminosity, X-ray detectability, and the possible origin of such strong AGN jets in the context of galaxy evolution. We conclude that end-on CR-dominated AGN bubbles could be a plausible scenario for the formation of ORCs.

\end{abstract}

\keywords{cosmic rays -- galaxies: active -- galaxies: clusters: intracluster medium -- hydrodynamics -- methods: numerical} 

\section{Introduction} \label{sec:intro}

Odd radio circles (ORCs) are extragalactic, edge-brightened, circular, low surface brightness radio objects that have no obvious counterpart emission in the optical, near-infrared (NIR), and X-ray bands. They were discovered in 2021 by \citet{Norris2021PASAORC} in the Australian SKA Pathfinder Telescope (ASKAP) EMU Pilot survey \citep[][]{Norris2021PASAPilot} and GMRT, with subsequent discoveries published by \citet{Koribalski2021MNRAS}, \citet{Koribalski2023arXiv230411784K}, and \citet{Gupta2022PASA}. Similar objects are also published by \citet{Filipovic2022MNRAS,Lochner2023MNRAS,Kumari2024MNRAS,Kumari2024A&A,Bulbul2024A&A,Koribalski2024MNRAS}. Optical/NIR spectroscopic follow-up observations of their suspected host galaxies have also been presented, where most of the suspected host galaxies are found to be quiescent massive ellipticals \citep{Rupke2023arXiv2IOP}. Moreover, one of them (ORC4) was found to have strong OII emission, suggesting violent activities happened in the past \citep{Coil2023arXiv2Nature}. 

After their discovery, several possible scenarios about their origin have been proposed, including structure formation shocks \citep{Yamasaki2023arXiv230917451Y}, AGN outflow driven shocks \citep{Fujita2023arXiv}, galaxy mergers \citep{Dolag2023ApJ}, or 
supernova remnants in the local group \citep[e.g.][]{Filipovic2022MNRAS,Sarbadhicary2023MNRAS}. 

\citet{Norris2022MNRAS} investigated one of the ORCs (ORC1) using the MeerKAT telescope \citep{Jonas2016mks_MeerKAT} and measured the spectral index and polarization with high angular resolution. They discussed three possible formation scenarios of the ORCs, including shocks from explosive events in the host galaxy, star formation termination shocks, and end-on radio bubbles generated by active galactic nucleus (AGN) jets. In the last scenario, relativistic jets from the central AGN would propagate into the intracluster/intragroup medium (ICM/IGM) and create low density regions (often referred to as radio bubbles or AGN bubbles) that are filled with cosmic rays (CRs) \citep[e.g.,][]{Dunn04}. The CRs within the bubbles could emit synchrotron radiation that lasts longer than the jet duration, depending on the cooling times of the CRs.
If the system is viewed side-on (that is, if the jet axis is perpendicular to the line of sight), it would be classified as typical radio bubbles. On the other hand, if the jet axis is nearly parallel to the line of sight, the radio-emitting bubbles could overlap with each other, forming a circular radio object that might be classified as an ORC. 

Previous AGN jet simulations \citep[see e.g.][for recent reviews]{Blandford2019,BourneYang2023Galaxy} 
rarely produce radio lobes or bubbles that are sufficiently large to explain the ORCs. 
For instance, in the simulations of \citet{Lin2023MNRAS}, the jets with power of $5\times 10^{45}$ erg s$^{-1}$ and duration of 10 Myr could only create bubbles with a radius of $ \sim 20$ kpc in a Perseus-like cluster environment (a cool-core cluster with virial mass $M_{\rm vir}\sim 8\times 10^{14}~M_\odot$), which is an order of magnitude smaller than what is required to explain the ORCs. There are potentially several ways to enlarge the size of the bubbles. After injection, the bubbles would gradually expand and approach pressure equilibrium with the ambient medium. Therefore, the size of the bubbles depends on the total energy contained within them and the ambient gas pressure profiles. 
The simplest way to expand the bubbles is thus to increase the power and duration of AGN jets. 
Another way is to decrease the mass of the system, so that the bubbles could expand more easily due to the lower pressure and weaker gravitational potential.

In this work, we aim to demonstrate that, under suitable parameter combinations, end-on AGN jet-inflated bubbles can reproduce properties of the observed ORCs.
We organize the structure of this paper as follows. In Section \ref{sec:method}, we describe the simulation setup and the treatment of CR transport and heating/cooling processes. In Section \ref{sec:result}, we describe the dynamics and expected radio morphology of the simulated bubbles. In Section \ref{sec:discussion}, we discuss the absolute brightness of the simulated ORCs, the possible origin of the strong jets in the context of galaxy evolution, and the caveats that should be addressed in future investigations. We summarize our conclusions in Section \ref{sec:conclusion}.

\section{Methods} \label{sec:method}
We perform three-dimensional (3D) CR-magnetohydrodyanmic (CR-MHD) simulations of AGN jets in a self-similar cluster environment using the adaptive mesh refinement (AMR) FLASH code \citep[][]{Flash}. The simulation setup and cluster environment are described in Section \ref{sec:simsetup} and Section \ref{sec:cluster}, respectively. We then produce simulated radio images by calculating synchrotron emissivity of the grid cells and integrating them along any given line of sight using methods described in Section \ref{sec:mockmethod}.

\subsection{CR-MHD simulations}\label{sec:simsetup}
In this work, we model the evolution of AGN jet-inflated bubbles by injecting mass, momentum, and energy of bipolar jets from a cylinder located at the center of the simulation domain, following prescriptions of \citet{Yang19} \citep[see also][]{Yang12a}. We assume the AGN jets are dominated by CR protons (CRp), identical to the CRpS simulation performed by \citet{Lin2023MNRAS} (S here represents the inclusion of CR streaming). As will be discussed in Section \ref{sec:result}, choosing CRp dominated jets is key for reproducing the observed features of ORCs. The simulation box size is set to 1 Mpc in order to simulate bubbles with a diameter over 500 kpc (similar to ORC1). The simulation time is set to 200 Myr.

Since the bubbles occupy a significant portion of the simulation box, and it would be too computationally expensive to refine the whole bubbles to maximum resolution using standard AMR criteria, we adopt the following strategy to determine the refinement levels within the simulation domain.
Specifically, we set the maximum refinement level to 9 within the central 20 kpc in order to properly resolve the jet launching region (with a maximum spatial resolution of 0.5 kpc). Regions located more than 20 kpc away from the center are assigned a maximum refinement level of 6, corresponding to a spatial resolution of 4 kpc. This approach significantly reduces the computational cost while still capturing the large-scale evolution of the bubbles. By performing convergence tests with different outskirt refinement levels, we verified that the simulated bubble size is robust.

We adopt the CR-MHD formalism \citep{Zweibel13, Zweibel17} to model the interaction between the injected CRs and thermal gas in the cluster. To first order, the CRs are coupled with the thermal gas because CRs are well scattered by tangled magnetic field. Therefore, on macroscopic scales (much greater than the CR gyro radii), one can treat CRs as a second fluid and solve the following CR-MHD equations:
\begin{equation}
\frac{\partial \rho}{\partial t}+\nabla \cdot(\rho \mathbf{v})=0
\end{equation}

\begin{equation}
\frac{\partial \rho \mathbf{v}}{\partial t}+\nabla \cdot\left(\rho \mathbf{v} \mathbf{v}-\frac{\mathbf{B} \mathbf{B}}{4 \pi}\right)+\nabla p_{\mathrm{tot}}=\rho \mathbf{g}
\end{equation}

\begin{equation}
\frac{\partial \mathbf{B}}{\partial t}-\nabla \times(\mathbf{v} \times \mathbf{B})=0
\end{equation}

\begin{eqnarray}
    \frac{\partial e}{\partial t}+\nabla \cdot\left[\left(e+p_{\mathrm{tot}}\right) \mathbf{v}-\frac{\mathbf{B}(\mathbf{B} \cdot \mathbf{v})}{4 \pi}\right]\nonumber\\
    =\rho \mathbf{v} \cdot \mathbf{g}+\nabla \cdot\left(\boldsymbol{\kappa} \cdot \nabla e_{\mathrm{cr}}\right) + \mathcal{H}_{\mathrm{cr}} - n_e^2\Lambda(T)
\end{eqnarray}

\begin{equation}
\frac{\partial e_{\mathrm{cr}}}{\partial t}+\nabla \cdot\left(e_{\mathrm{cr}} \mathbf{v}\right)=-p_{\mathrm{cr}} \nabla \cdot \mathbf{v}+\nabla \cdot\left(\boldsymbol{\kappa} \cdot \nabla e_{\mathrm{cr}}\right) + \mathcal{C}_{\mathrm{cr}}
\end{equation}
where $\rho$ and $\mathbf{v}$ are the gas density and velocity, respectively, $\mathbf{g}$ is the gravitational field, and $\boldsymbol{\kappa}$ is the CR diffusion tensor, which describes the second-order effect of CR transport. $e_{\textrm{cr}}$ is the CR energy density and $e$ is the total energy density, consisting of kinetic, thermal, CR and magnetic energy ($e=0.5\rho v^2 + e_{\textrm{th}} + e_{\textrm{cr}} + B^2/8\pi$). The total pressure is $p_{\textrm{tot}} = (\gamma-1)e_{\textrm{th}} + (\gamma_{\textrm{cr}}-1)e_{\textrm{cr}} + B^2/8\pi$, in which we adopt the adiabatic index $\gamma=5/3$ for the thermal gas and $\gamma_{\textrm{cr}}=4/3$ for relativistic CRs.
$\mathcal{H}_{\mathrm{cr}}$ is the net contribution of CR related processes to the change of total energy density, $\mathcal{C}_{\mathrm{cr}}$ is the CR cooling rate due to the combined effect of Coulomb losses, hadronic processes, and CR streaming. Inverse-Compton (IC) scattering and synchrotron losses are neglected in the simulations because the jets are assumed to be CRp dominated. $n_e$ is the electron number density, and $\Lambda(T)$ is the radiative cooling function. We refer the readers to \citet{Lin2023MNRAS} for additional details.

\subsection{Setup of the cluster environment}\label{sec:cluster}
We assume that the gas in the cluster follows the same temperature profile as the observed Perseus cluster and is initially in hydrostatic equilibrium within a static gravitational potential described by the NFW profile \citep[][]{NFW}. The system is characterized by the virial mass ($M_{\rm vir}$), virial radius ($R_{\rm vir}$), central temperature ($T_c$), concentration parameter ($c$) and baryon fraction ($M_g/M_{\rm vir}$).
\begin{table}[]
    \centering
    \begin{tabular}{|c|c|c|}
        \hline
        Parameter & Symbol & Value\\
        \hline
        Virial Mass & $M_{\rm vir}$ & $8\times 10^{14}~M_\odot$ \\
        Virial Radius & $R_{\rm vir}$ & 2440 kpc \\
        Baryon fraction & $M_g/M_{\rm vir}$ & 0.15 \\ 
        Central temperature & $T_c$ & 7 keV \\
        Concentration & $c$ & 6.81\\
        Magnetic field normalization& $b_{\rm norm}$ & 0.01\\
        Jet power & $P_0$ & $5\times10^{45}$ erg s$^{-1}$ \\
        Jet duration & $D_0$ & 10 Myr \\
        \hline
    \end{tabular}
    \caption{Fiducial parameters for our cluster setup.}
    \label{tab:fiducial_cluster}
\end{table}
As mentioned in Section \ref{sec:intro}, to produce bubbles with a radius comparable to the observed ORCs, a smaller ambient pressure, and thus a less massive cluster is favored. To this end, we set up the gas profiles using the self-similar scaling relations of galaxy clusters \citep[][]{Kaiser86, Yang16a}:
\begin{equation}
    R_{\rm vir}\propto M_\mathrm{vir}^{1/3}, \quad T_c\propto M_\mathrm{vir}^{2/3}, \quad c\propto M_\mathrm{vir}^{-0.08}.
\end{equation}
A tangled magnetic field is generated by performing 3D inverse Fourier transform of a magnetic power spectrum with a coherence length of 100 kpc. Following the notation in \citet{Yang12a}, the slope of the magnetic power spectrum in Fourier space is $-11/6$, and the cutoff wavenumbers of the spectrum corresponds to physical scales of 90 kpc and 210 kpc. We refer the readers to \citet{Yang12a}, \citet{Yang13}, and \citet{Yang16a} for details. To fix the plasma beta parameter $\beta=P_{\rm th}/P_{\rm B}$, we normalize the magnetic field strength with a parameter $b_{\rm norm}\propto M_{\rm vir}^{2/3}$ to keep a constant value of $\beta \sim 100$ \citep{Carilli02, Feretti2012A&ARv}. The fiducial parameters are summarized in Table \ref{tab:fiducial_cluster}.

For the following discussion, we employ the naming convention [Physics]\_[Mass]\_[Power]\_[Duration] in our simulations. For instance, the simulation denoted as CRpS\_M12\_P5\_D5 represents the simulation that uses CRpS jets in a cluster with virial mass $M_{\rm vir}=8\times 10^{12}~M_\odot$, jet power of five times the fiducial value ($2.5\times 10^{46} ~{\rm erg~s^{-1}}$), and jets that last for five times the fiducial duration (50 Myr).

\subsection{Radio signatures}\label{sec:mockmethod}
In the hadronic scenario considered in this study, the radio emission comes from synchrotron radiation of secondary electrons and positrons produced by the hadronic collisions between the injected CRp and the ambient medium \citep[e.g.,][]{Owen2022MNRAS}. While it is possible to calculate the synchrotron emissivity \citep[using methods described in Appendix C in][]{Lin2023MNRAS}, solving for the distributions of secondary particles across the whole simulation box is computationally expensive and unnecessary for the purpose of simple morphological comparison. Also, to compute the absolute values of the synchrotron emissivity involves a number of uncertainties (see discussion in Section \ref{sec:discussion-AB}). Therefore, in this study, we adopt a simplified method that assumes the radio emissivity of a given cell to be proportional to the hadronic cooling rate of CRp (which approximates the secondary-particle production rate under the steady-state assumption) and the strength of the magnetic field squared \citep[$\epsilon \propto \mathcal{C}_{\mathrm{CRp}, \mathrm{h}} \propto e_{\rm cr}\rho B^2$, ][]{YoastHull13,Ruszkowski17}. 
After calculating the emissivity of each grid cell, we then perform integration along a given axis and obtain a predicted radio image.

To allow direct morphological comparison to the observed ORCs, we convolve the radio image with a Gaussian kernel that has a standard deviation ($\sigma$) corresponding to a 6-arcsecond beam size \citep{Norris2022MNRAS}. The conversion of the kernel size between the angular size (in units of arcseconds) and physical size (in units of pixels on the image) is done by considering the angular diameter distance ($d_A(z)$) in a flat LambdaCDM cosmology with $\Omega_{\rm m,0}=0.3$ and $h_0=0.7$, i.e.,  
\begin{eqnarray}
    \sigma ({\rm pixel}) = 25.67\left( \frac{d_A(z)}{\rm 1324~Mpc} \right) \left( \frac{\rm Beam~size}{\rm 6~arcsecond} \right)\nonumber\\
    \times \left( \frac{\rm Image~ size}{\rm 800~pixel} \right) \left( \frac{\rm Box~ size}{\rm 1200~ kpc} \right),
\end{eqnarray}
where $d_A(z)$ depends on the redshift ($z$) of the ORC of interest and is calculated using the astropy.cosmology Python module \citep{Astropy2022ApJ}.

\section{Results} \label{sec:result}
\subsection{Benchmark cases}\label{sec:benchmark}
In this section, we describe two of the most promising simulations, namely CRpS\_M12\_P5\_D5 and CRpS\_M13\_P5\_D5 in detail. 

\begin{figure}[htbp]
    \centering
    \includegraphics[width=\columnwidth]{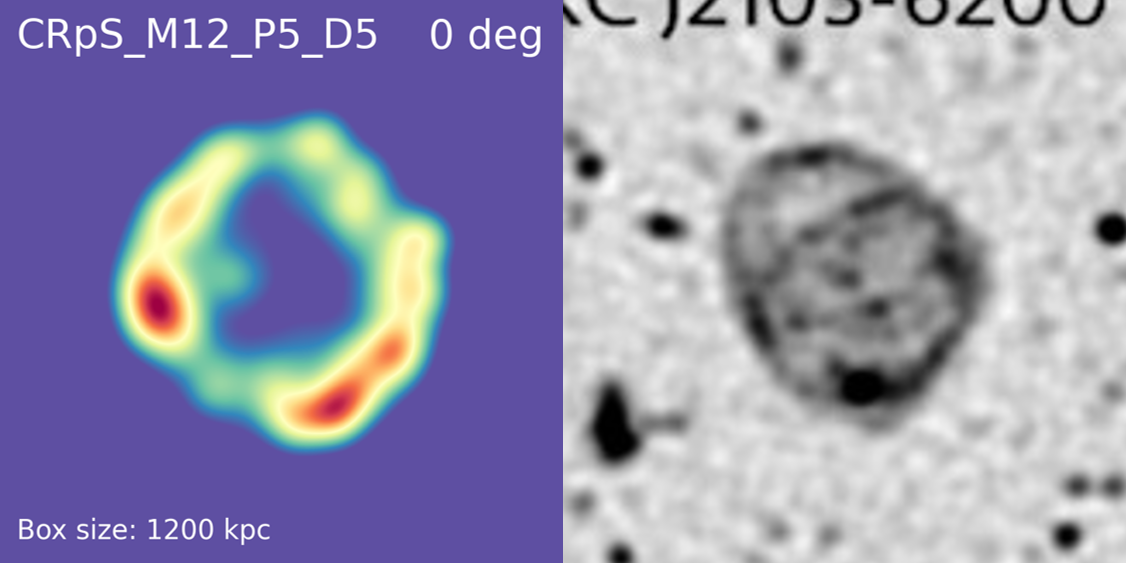}
    \includegraphics[width=\columnwidth]{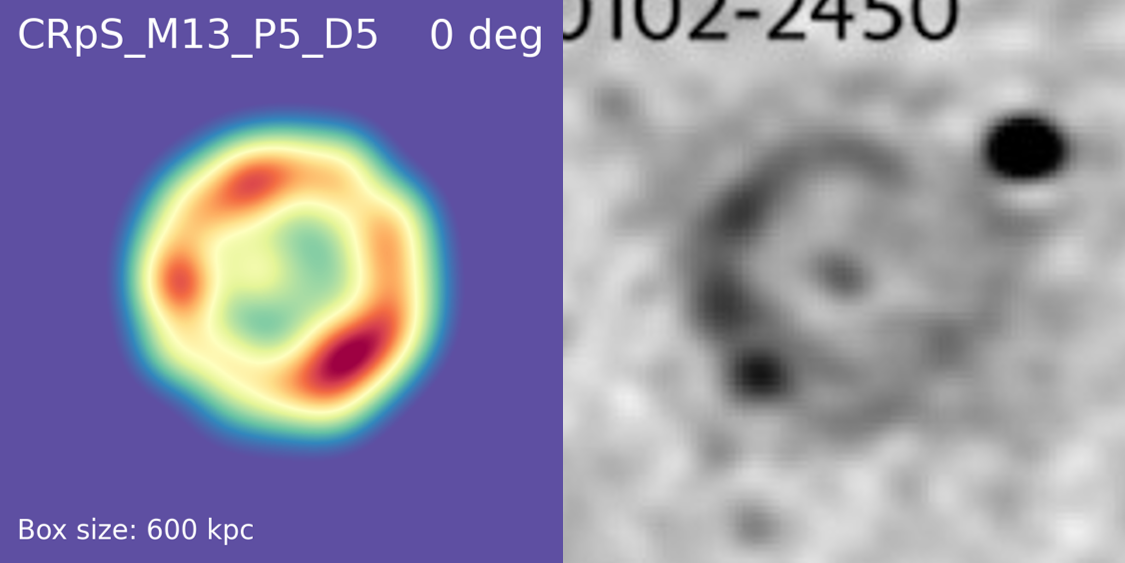}
    \caption{{\it Top panel:} Comparison between the simulated radio image in CRpS\_M12\_P5\_D5 at 200 Myr (left) and the observed ORC1 (right). {\it Bottom panel:} Comparison between the images of CRpS\_M12\_P5\_D5 at 200 Myr and ORC5. Images of ORC1 and ORC5 (observed at 1284 and 944 MHz, respectively) are retrieved from \citet{Norris2022MNRAS} and cut to approximately the same physical size as the simulated ones. The simulated images are obtained by assuming a viewing angle aligned with the jet axis and are smoothed by assuming that they are located at $z=0.551$ (same as ORC1) and $z=0.27$ (same as ORC5), respectively. The colorscale is linearly normalized to the maximum brightness in each image.}
    \label{fig:ORC_morphology}
\end{figure}

The top panel of Fig. \ref{fig:ORC_morphology} displays the simulated radio image of CRpS\_M12\_P5\_D5 (left) at 200 Myr with a viewing angle of $0^\circ$ (line of sight parallel to the jet axis) and the observed image of ORC1 (right), linearly normalized with the maximum value in the image. Both images have a side length of roughly 1200 kpc. From the images, it is evident that end-on AGN jet-inflated bubbles can indeed produce a circular radio object with a diameter comparable to ORC1 \citep[520 kpc,][]{Norris2022MNRAS}. The edge-brightened feature is also reproduced with reasonable contrast between the edge and the interior of the object. Consistencies can also be found in the comparison between CRpS\_M13\_P5\_D5 and ORC5 (bottom panel of Fig. \ref{fig:ORC_morphology}; each image has a side length of 600 kpc). Note that due to the deeper gravitational potential well and higher ambient gas pressure, the bubble radius in CRpS\_M13\_P5\_D5 is roughly half the size than that in CRpS\_M12\_P5\_D5, despite the same jet power and duration. Furthermore, because of the large difference in bubble volume, the surface brightness of CRpS\_M12\_P5\_D5 is roughly 1000 times fainter than CRpS\_M13\_P5\_D5 since its CR energy density ($e_{\rm cr}$), gas density ($\rho$), and magnetic field strength ($B$) are all smaller in the lower mass system.

\begin{figure}[htbp]
    \centering
    \includegraphics[width=\columnwidth]{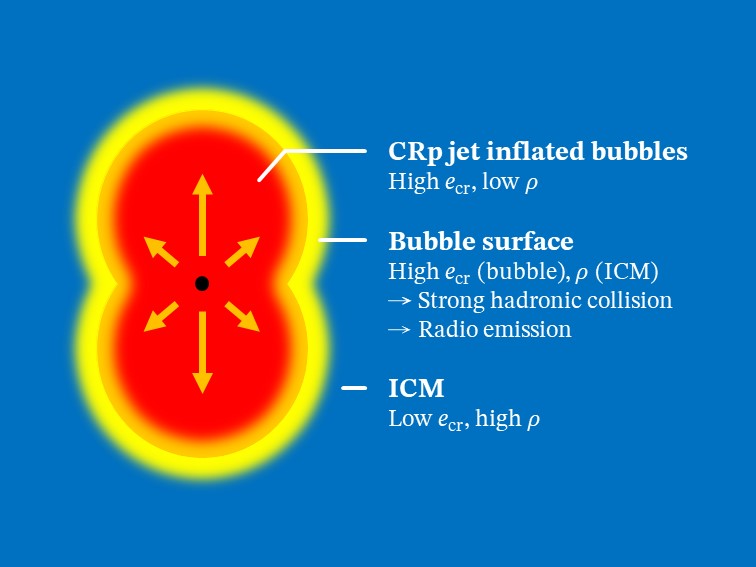}
    \caption{Schematic diagram that illustrate the key idea of the hadronic AGN bubble scenario. By combining the high gas density in the ambient ICM and high CRp energy density within the bubbles, the surface of the bubbles can naturally reproduce the edge-brightened feature of the observed ORCs after projected along the jet axis.}
    \label{fig:infographic-KeyIdea}
\end{figure}

The key mechanism that enables us to reproduce the limb-brightened feature lies in the nature of the hadronic processes. As mentioned in Section \ref{sec:mockmethod}, the synchrotron emissivity is directly proportional to $e_{\rm cr}$ and $\rho$. Although $e_{\rm cr}$ is high within the bubbles, hadronic collisions are inefficient due to the low gas density. Conversely, the ambient ICM exhibits a relatively high gas density but lacks CRs for interaction. The confluence of high $e_{\rm cr}$ and $\rho$ occurs only at the interface between the bubbles and ICM. A schematic diagram illustrating this idea is shown in Fig.\ \ref{fig:infographic-KeyIdea}.

\begin{figure*}[htbp]
    \centering
    \includegraphics[width=\textwidth]{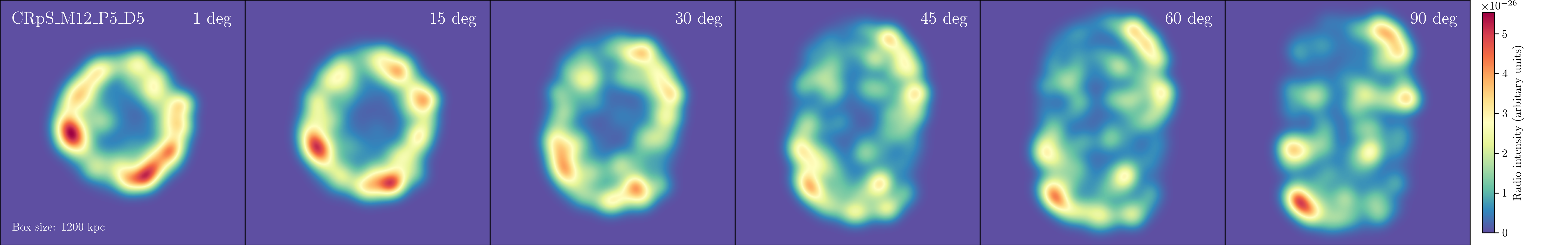}
    \includegraphics[width=\textwidth]{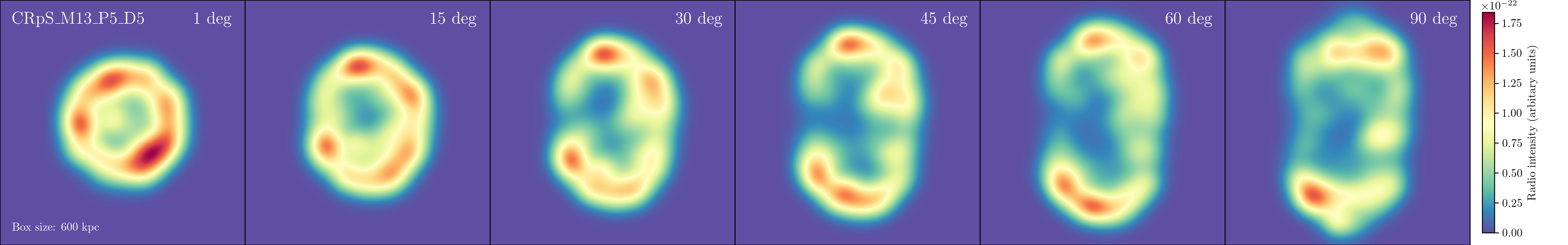}
    \caption{Simulated radio images of CRpS\_M12\_P5\_D5 and CRpS\_M13\_P5\_D5 at 200 Myr with viewing angles ranging from $0^\circ$ to $90^\circ$ with repsect to the jet axis. The box sizes are 1200 kpc (top row) and 600 kpc (bottom row), respectively. While the absolute value of the colorbar is arbitrary, the relative value between the two simulations does carry physical meanings -- the ORC in CRpS\_M12\_P5\_D5 is significantly fainter than that in CRpS\_M13\_P5\_D5 (see discussion in Section \ref{sec:benchmark}). 
    }
    \label{fig:ORC_viewingangle}
\end{figure*}

One drawback of the AGN scenario, as discussed in \citet{Norris2022MNRAS}, is the requirement of a highly aligned jet axis with the line of sight, which seems coincidental. 
To investigate the dependence of ORC properties on the viewing angle, we present the radio maps of both CRpS\_M12\_P5\_D5 and CRpS\_M13\_P5\_D5 at 200 Myr with viewing angles ranging from $0^\circ$ to $90^\circ$ in Fig. \ref{fig:ORC_viewingangle}. It is evident that the eccentricity of the rings remains consistent with ORC1 up to an angle of $\sim 30^\circ$ between the jet axis and the line of sight. The shape of the ring is relatively insensitive to the viewing angle because CR dominated jets tend to generate round bubbles \citep[rather than elongated bubbles inflated by kinetic-energy dominated jets;][]{Yang19}. Although predicting the probabilities of observing the ORCs and their number densities requires further investigation, our results suggest that the issue of viewing-angle alignment may not be as severe as one may naively expect. 

It is also worth noting that the emission of the simulated bubbles is not smoothly distributed at the interface between the bubbles and the ICM. Instead, a significant portion of the emission originates from numerous small, luminous clumps scattered across the surface of the bubbles, which are formed due to local fluid and thermal instabilities. As the gas cools and condenses into these dense clumps, not only does the hadronic collision rate increase, but the magnetic field strength is also enhanced due to flux freezing. The ring-like structure in Fig. \ref{fig:ORC_morphology} actually consists of multiple small and bright synchrotron-emitting clumps rather than a single, smooth surface. Therefore, as the viewing angle increases, the system would become very clumpy and may not be classified as an ORC observationally (e.g. CRpS\_M12\_P5\_D5 with $90^\circ$ viewing angle in Fig. \ref{fig:ORC_viewingangle}). Moreover, our results imply that with higher resolution observations (e.g. VLA or SKA), one should be able to resolve the rings into many small and bright clumps. This feature could serve as one prediction of our model. 

\subsection{Parameter search}\label{sec:parasearch}
\begin{figure*}[htbp]
    \centering
    \includegraphics[width=\textwidth]{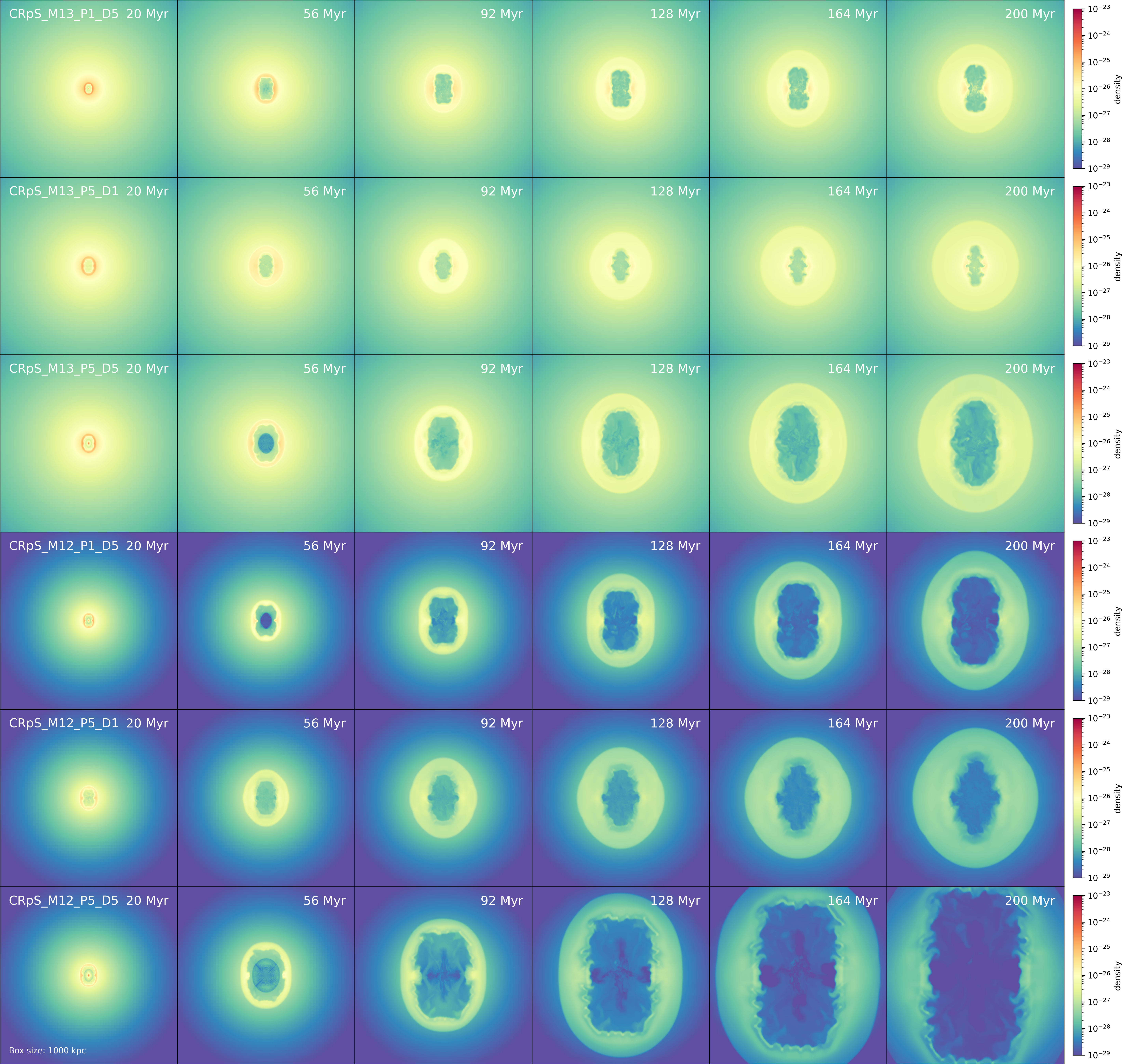}
    \caption{Time series of the side-on density slices of the six simulations described in Section \ref{sec:parasearch}. The top three rows show systems with virial mass $M_{\rm vir}=8\times 10^{13}~M_\odot$, and the bottom three rows are for cluster masses of $M_{\rm tot}=8\times 10^{12}~M_\odot$. For each cluster mass, the rows from top to bottom show simulations with the jet power and duration of $(P_0, 5D_0)$, $(5P_0, D_0)$, and $(5P_0, 5D_0)$.}
    \label{fig:ORC_timeseries}
\end{figure*}

\begin{figure*}[htbp]
    \centering
    \includegraphics[width=\textwidth]{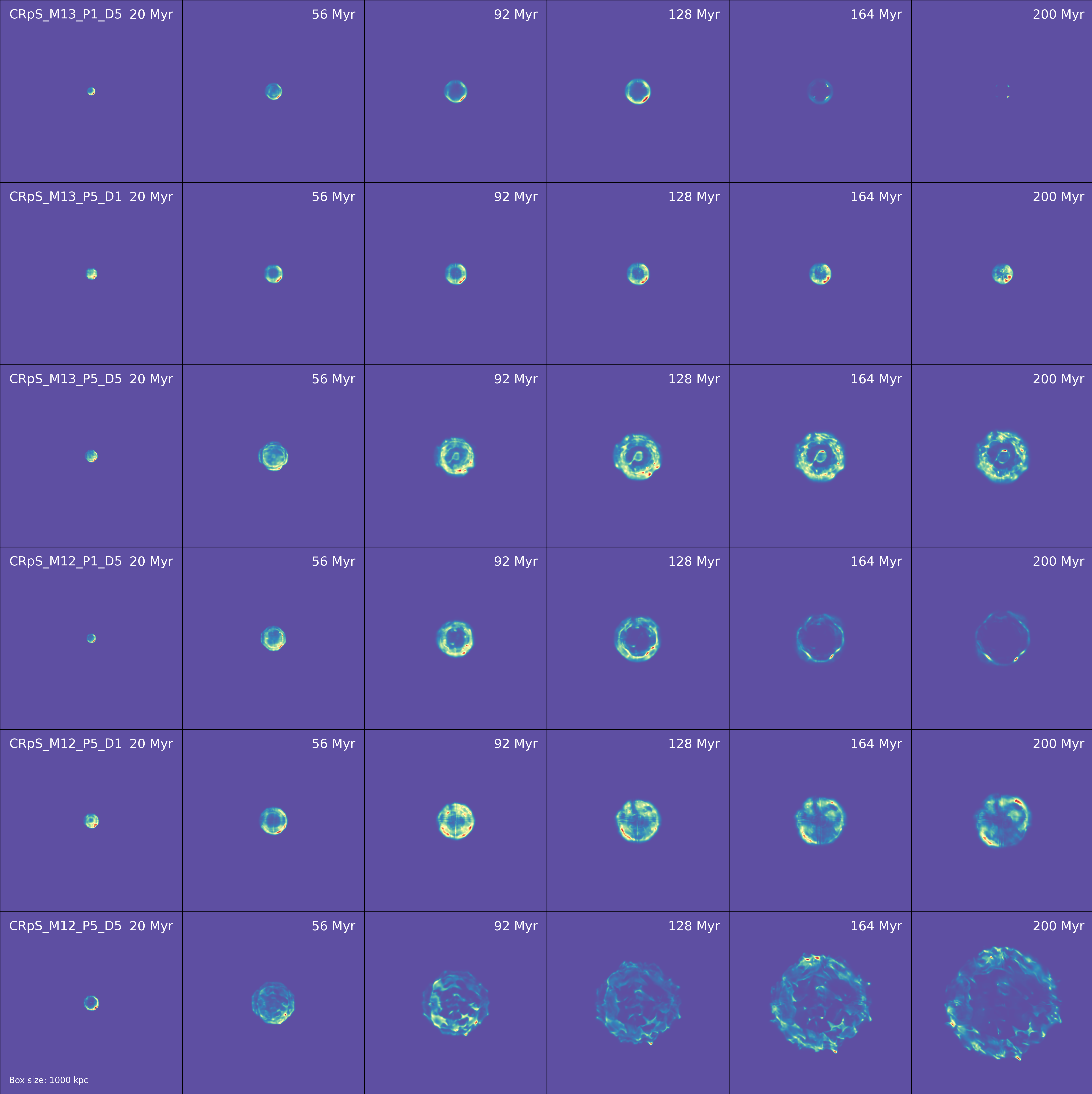}
    \caption{Time series of the radio morphology of the six simulations described in Section \ref{sec:parasearch}. The order of each panel is the same as in Fig. \ref{fig:ORC_timeseries}.  Note that the surface brightness in each image is normalized by its maximum value in order to emphasize the morphology. In reality, synchrotron emission from earlier snapshots are generally much brighter than the later ones since the CR energy density deceases due to adiabatic expansion.}
    \label{fig:ORC_emissiontimeseries}
\end{figure*}

\begin{figure*}[htbp]
    \centering
    \includegraphics[width=\textwidth]{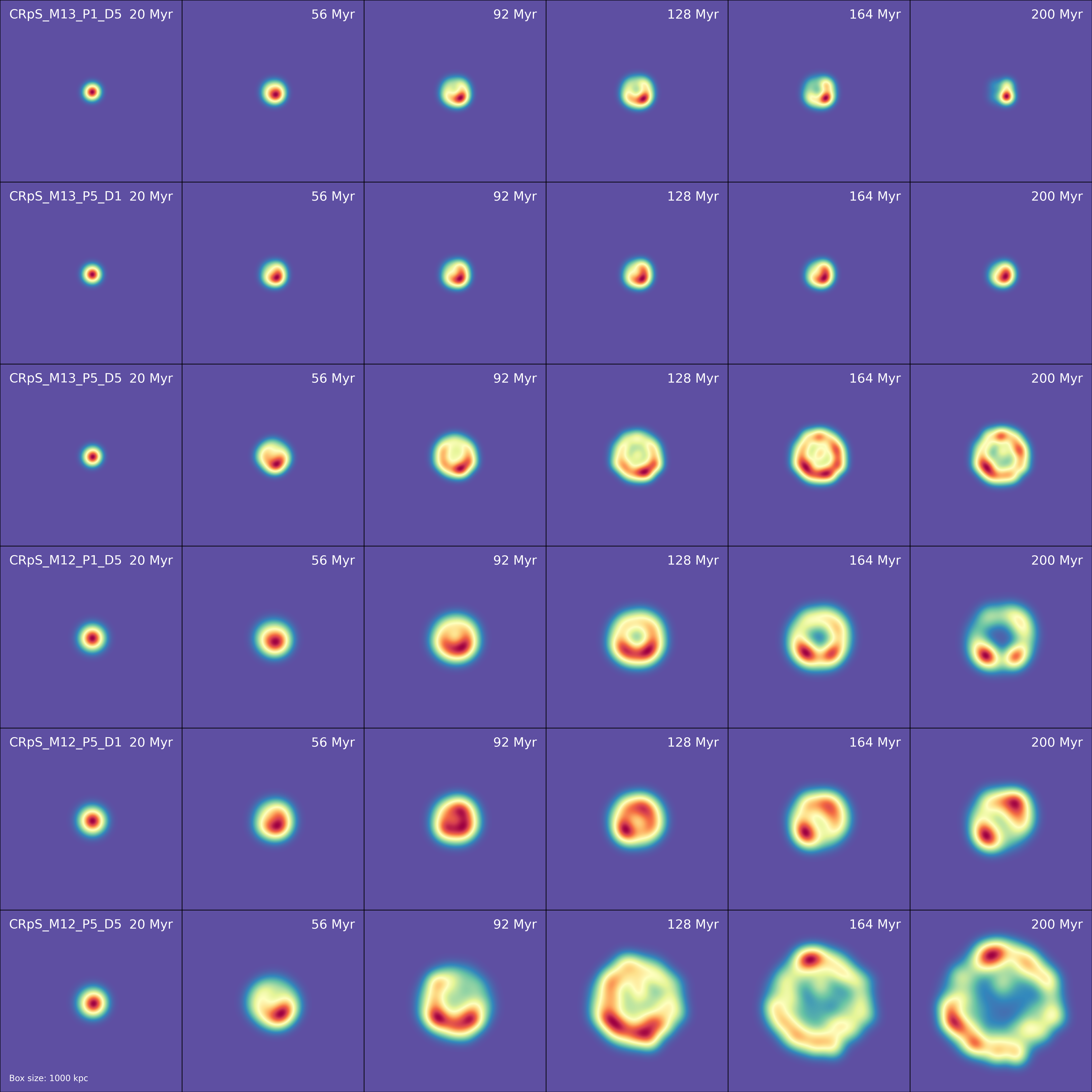}
    \caption{As Fig. \ref{fig:ORC_emissiontimeseries} but smoothed with the 6" beam size. M13 ($M_{\rm vir}=8\times 10^{13} M_\odot$) simulations are assumed to be located at $z=0.27$ (i.e., the observed redshift of ORC5) and M12 ($M_{\rm vir}=8\times 10^{12} M_\odot$) simulations are assumed to be located at $z=0.551$ (the redshift of ORC1).}
    \label{fig:ORC_emissiontimeseries_smoothed}
\end{figure*}

To explore the parameter dependence of the simulated ORCs, we carry out a series of simulations to investigate how the jet power, jet duration, and cluster mass affect the dynamics and radio morphology of the bubbles. Fig. \ref{fig:ORC_timeseries} shows the side-on time series of density slices of simulations with different parameter combinations. The top three rows show the results for a virial mass $M_{\rm vir}=8\times 10^{13}~M_\odot$, and the bottom three rows are for $M_{\rm vir}=8\times 10^{12}~M_\odot$. Among the simulations with the same cluster mass, the first row represents a simulation with the fiducial jet power ($P_0$) but a long jet duration ($5D_0$). The second row shows the results for a high jet power ($5P_0$) and the fiducial jet duration ($D_0$). The simulation shown in the third row, as described in Section \ref{sec:benchmark}, features both a high jet power ($5P_0$) and a high jet duration ($5D_0$). In Fig. \ref{fig:ORC_emissiontimeseries} and \ref{fig:ORC_emissiontimeseries_smoothed}, we show the intrinsic and smoothed end-on radio images of the corresponding simulations, respectively. Several points can be drawn from the figures:
\begin{enumerate}
    \item Given the same total input energy, systems with a higher jet power tend to produce bubbles with a smaller volume. This is because more powerful jets would generate stronger shocks that carry energy away and reduce the amount of energy stored within the bubbles to drive bubble expansion. This is consistent with previous findings by \citet{Tang17}. 
    \item If the total input energy is insufficient, the ICM would efficiently cool via radiative cooling and collapse back onto the bubbles, as seen in the first and second rows in Fig. \ref{fig:ORC_timeseries}. By contrast, in the low mass ($M_{\rm vir}=8\times 10^{12}~M_\odot$) system, the same jets can evacuate the ambient medium from the central region more effectively. 
    \item As mentioned in Section \ref{sec:result}, the radio emission in our simulations mainly comes from small and bright clumps formed via hydrodynamic and thermal instabilities near the bubble surface. Therefore, the beam size of the observation would have a great impact on the morphology of the objects. For instance, with a 6-arcsecond beam, CRpS\_M13\_P5\_D5 and CRpS\_M12\_P1\_D5 exhibit very similar radio morphology (Fig. \ref{fig:ORC_emissiontimeseries_smoothed}). However, their intrinsic morphology (e.g. the number of bright clumps and brightness contrast between their edges and interiors) are significantly different (Fig. \ref{fig:ORC_emissiontimeseries}). The underlying physical properties of the two systems (cluster mass and jet power) are also substantially different.
\end{enumerate}

It is worth noting that the physical picture described above -- wherein long duration jets and lower ambient pressure favor the formation of larger bubbles -- differs from many radio galaxy simulations \citep[e.g.][]{Norman1983ASSL,Gaibler2009MNRAS,English2016MNRAS}. The different trends could be owing to the fact that most radio galaxy simulations use relativistically moving jets with $v\sim c$, which are kinetic energy dominated \citep[see e.g.][for a recent example]{Stimpson2023MNRAS}, and hence they have sufficient momentum to remain collimated and (relatively) weakly coupled to the ambient gas. In contrast, our jets are CR dominated with $\gamma \sim 4/3$ and sub-relativistic bulk velocities ($v_{\rm inj} \sim 0.1 c$). Therefore, the bubble expansion is driven primarily by pressure contrast instead of momentum, resulting in bubbles that are more prone to pressure confinement of the ambient medium.

\section{Discussion}\label{sec:discussion}
\subsection{Requirements for reproducing the ORCs}\label{sec:hadronic}
In this section, we discuss several lessons learned for creating ORC-like objects using CRp dominated AGN bubbles. 

First of all, in order to produce bubbles with radii comparable to the observed ORCs, the power and duration of the jets must be strong enough to completely evacuate the high density gas originally sitting near the center of the cluster. If this is not achieved, the central high density gas would easily dominate the synchrotron emission ($\epsilon \propto e_{\rm cr} \rho B^2$) and create a centrally brightened radio source. Note that even if the ICM is temporarily evacuated, gas could still cool (especially if compressed) through radiative cooling and eventually fall back toward the central region within tens of Myr (see e.g. CRpS\_M13\_P1\_D5 at 200 Myr in Fig. \ref{fig:ORC_timeseries}). Therefore, the total energy input from the AGN jets should be significantly larger than the gravitational potential energy of the inner ICM/IGM.

CRpS jets could naturally reproduce the edge-brightened morphology of the observed ORCs because the hadronic collision is most efficient at the interface between the bubbles and the ICM (Fig \ref{fig:infographic-KeyIdea}). On the other hand, if the jets are CRe dominated \citep[][]{Dunn04, Birzan08, Croston18, Lin2023MNRAS}, the synchrotron emissivity is instead $\epsilon\propto e_{\rm cr}B^2$. In this case, one would generally produce a centrally brightened radio object instead of an edge-brightened one, because the synchrotron emissivity is largely uniform in the bubble interior, and the pathlength is the greatest along the jet axis. The edge-brightened feature could be produced in some cases if the magnetic field strength is enhanced by adiabatic compression at the interface between the bubbles and the ICM, where the compression could occur due to bubble expansion and fallback of the ICM in response to radiative cooling. However, this feature is short-lived and requires fine-tuning of the parameters. Most important of all, given the old age (200 Myr) of the bubbles, the primary CRe injected by the AGNs should have cooled below the energy required to shine at the observed frequency. For instance, in \citet{Norris2022MNRAS}, ORC1 ($z=0.551$) is observed with MeerKAT at around 1 GHz, we can therefore estimate the corresponding CRe energy to be:

\begin{equation}
    E_{\rm CRe}(\nu_{\rm obs}) = \gamma m_e c^2 \sim \sqrt{\frac{(1+z)\nu_{\rm obs}}{\nu_G}} m_e c^2,
\end{equation}
where $\gamma$ is the Lorentz factor of the CRe, and the gyro frequency $\nu_G$ can be obtained by the relation
\begin{equation}
    \frac{\nu_G}{\rm MHz} = 2.8\times \frac{B}{\rm Gauss}.
\end{equation}
Assuming $B=0.1~ \mu G$, the corresponding CRe energy is $E_{\rm CRe}({\rm 1~GHz})\sim 40 {\rm ~GeV}$. Because the magnetic field strength is small, CRe cooling is dominated by IC scattering of the cosmic microwave background (CMB) photons. We can then estimate the CRe cooling time with Eqn. 65 in \citet{Miniati2001}
\begin{equation}
\tau_{\rm IC} \sim 2.3 \times 10^{12} \gamma^{-1}(1+z)^{-4}~{\rm yr}
\end{equation}
For $z=0.551$ and $\gamma = E_{\rm CRe}/(m_e c^2) \sim 8\times 10^4$, the CRe cooling time is $\sim 5$ Myr, significantly shorter than the formation time of the bubbles. 
It is therefore evident that given the bubble age of 200 Myr, the primary CRe from AGN jets would have cooled below the observable frequency, necessitating shock/turbulence reacceleration in the leptonic scenario \citep{Shabala2024PASA}.


\subsection{Remarks on the absolute brightness}\label{sec:discussion-AB}
While our setup could successfully reproduce the size and morphology (that is, relative brightness) of the ORCs, it is important to ask whether the specific intensity produced by our simulated ORCs are comparable to the observed ones. To this end, we provide a simple order-of-magnitude estimation of the flux from our simulated ORCs. As will be discussed below, there are parameter and modeling uncertainties that could significantly affect our estimates. However, in order to see whether our proposed model is plausible at all, a crude estimate is still useful, which is described as follows.

Assuming that the total power of secondary CRe cooling, which includes IC scattering with CMB photons and synchrotron emission, is roughly equal to the energy injection rate from hadronic collisions, we can express the total luminosity of the system as (see Appendix \ref{sec:Fnu_estimate} for the detailed descriptions):

\begin{equation}
    \epsilon_{{\rm sync},\nu} = \frac{P_{\rm had}}{6}\frac{3-p}{2}\nu^{\frac{1-p}{2}}\left(\nu_{\max}^{\frac{3-p}{2}}-\nu_{\min}^{\frac{3-p}{2}}\right)^{-1}\frac{u_B}{u_B+u_{\rm CMB}}
\end{equation}

\begin{equation}\label{eq:F_nu}
    F_\nu = \frac{L_\nu}{4\pi d_{\rm L}^2(z)},\quad {\rm where~}L_{\nu} = \int \epsilon_{{\rm sync},\nu} dV.
\end{equation}

Here $P_{\rm had}$ is the hadronic heating rate (in units of erg s$^{-1}$) within each simulation grid cell, $p$ is the spectral index of secondary CRe ($n(E)\propto E^{-p}$), $\nu$ is the observed frequency, $\nu_{\min}$ and $\nu_{\max}$ are the frequencies of synchrotron emission corresponding to the cutoff of the CRe spectrum at low and high energies, respectively, $u_B$ and $u_{\rm CMB}$ are the energy densities of the magnetic field and CMB, and $d_{\rm L}(z)$ is the luminosity distance. Using $p\sim 2.6$ \citep[motivated by observations,][]{Koribalski2021MNRAS} and $E_{\max}\sim 100$ GeV (typically assumed in CR simulations) in Equation \ref{eq:F_nu}, we estimate that the flux at 1 GHz for the M12 and M13 (note that M12 represents $M_{\rm vir}=8\times 10^{12}~M_\odot$, and are placed at $z=0.551$ and $z=0.27$, respectively) simulations at $t=200$ Myr is around $10^{-7}$ to $10^{-3}$ mJy, respectively. Assuming a flatter CR spectrum (smaller $p$) and/or observing the simulated ORC at an earlier times ($t\sim 100$ Myr, when the CRp density and hence the hadronic collision rate is higher) would yield higher fluxes up to $\sim 10^{-2}$ mJy. These estimated fluxes fall short compared to those of the observed ORCs, which have fluxes on the order of mJy. However, the above estimate has only considered the secondary CRe that are {\it instantaneously} injected via hadronic collisions and neglected the accumulation and re-acceleration of these electrons on the bubble surface. Given these considerations, we therefore argue that the estimated values should be regarded as lower limits of the actual fluxes.

\begin{figure}[htbp]
    \centering
    \includegraphics[width=\columnwidth]{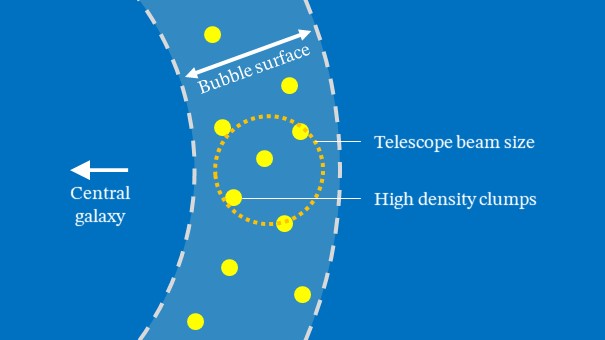}
    \caption{Schematic diagram of the structure of bubble surface.}
    \label{fig:infographic-AbsoluteLuminosity}
\end{figure}

There are several other factors that can cause large uncertainties in the calculations of the absolute brightness in our model. First of all, since the radio emission mainly comes from high density clumps originating from hydrodynamic instabilities and radiative cooling, the predicted intensity of the ORCs is very sensitive to how large the cold clumps are, which is typically non-convergent in numerical simulations. The dependence of observed intensity on clump size can be understood as follows. Consider a region on the bubble surface that is enclosed by a sphere with the radius of the observed beam size ($R_{\rm beam}$), as illustrated in Fig \ref{fig:infographic-AbsoluteLuminosity}, the total enclosed mass
\begin{equation}
    M_{\rm beam}= NR_c^3\rho_c
\end{equation}
is roughly a constant with varying simulation resolution, where $N$ is the number of clumps within the sphere, $R_c$ is the radius of the clump, and $\rho_c$ is the density of the clump. Therefore, the effective emissivity is the total luminosity of the clumps ($NR_c^3\epsilon_c$) divided by the volume of the sphere
\begin{equation}
    \epsilon_{\rm beam} = \frac{NR_c^3 \epsilon_{c}}{R_{\rm beam}^3} \propto \frac{NR_c^3\rho_c e_{\rm cr}B^2}{R_{\rm beam}^3} \propto e_{\rm cr}B^2.
\end{equation}
With increasing simulation resolution, the clumps would be better resolved (with reduced $R_c$), and $e_{\rm cr}$ and $B$ would both be enhanced in the clumps due to the compression. It is therefore hard to predict the exact brightness of the simulated ORCs. 
The assumed magnetic field strength also has a great impact on the synchrotron emissivity ($\epsilon_{\rm sync}\propto B^2$). While the plasma $\beta$ is observationally constrained to approximately 100 in galaxy clusters, there exists no definitive constraint on the magnetic field strength at the scale of hundreds of kpc for a group-like system. Using the self-similar scaling relations described in Section \ref{sec:method}, the magnetic field strength for the M12 and M13 halos at the scales of a few hundreds of kpc is on the order of 0.3 $\mu$G and 0.06 $\mu$G\footnote{This is obtained by doing an synchrotron emissivity weighted average of the magnetic field strength.}, respectively. A mere scaling of the magnetic field strength to a few $\mu \rm G$ could result in a remarkable increase in synchrotron emissivity by a factor of $10^4$. Given the lack of constraints on crucial factors on both theoretical (clump sizes) and observational (magnetic field strength) sides, we conclude that a robust prediction of the absolute luminosity should be addressed with more sophisticated simulations and is left for future investigation.\footnote{In fact, most of the theoretical studies of the ORCs so far have not yet presented clear conclusions on the absolute brightness.}

\subsection{Detectability in X-ray}\label{sec:mockXray}
To further investigate the possible observable signatures of the AGN bubble scenario, we use pyXSIM \citep{Biffi2012MNRAS,Biffi2013MNRAS,ZuHoneHallman2016ascl} to predict whether our simulated ORCs can be detected by current and future X-ray observatories, namely Chandra ACIS-I, AXIS \citep[][]{AXISOverview_Reynolds2023}, and the Wide Field Instrument (WFI) on Athena.\footnote{ESA’s Science Programme Committee has endorsed a rescoped version of the Athena X-ray observatory, called NewAthena, on November 8th, 2023. While the design of the mission is revised, the overall capability remains comparable.} Details regarding the mock observations can be found in Appendix \ref{sec:pyxsim_details}. 

\begin{figure}[htbp]
    \centering
    \includegraphics[width=\columnwidth]{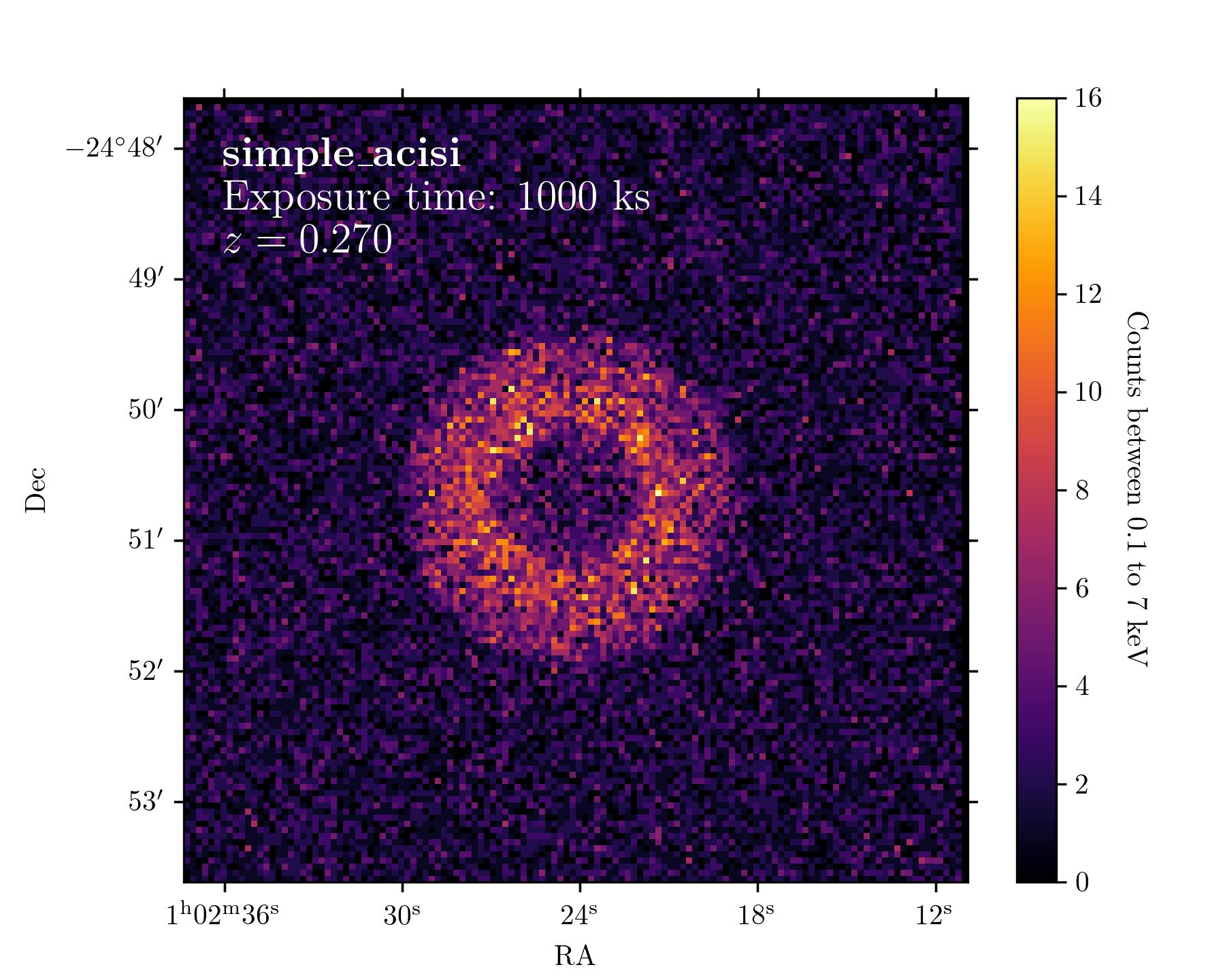}
    \includegraphics[width=\columnwidth]{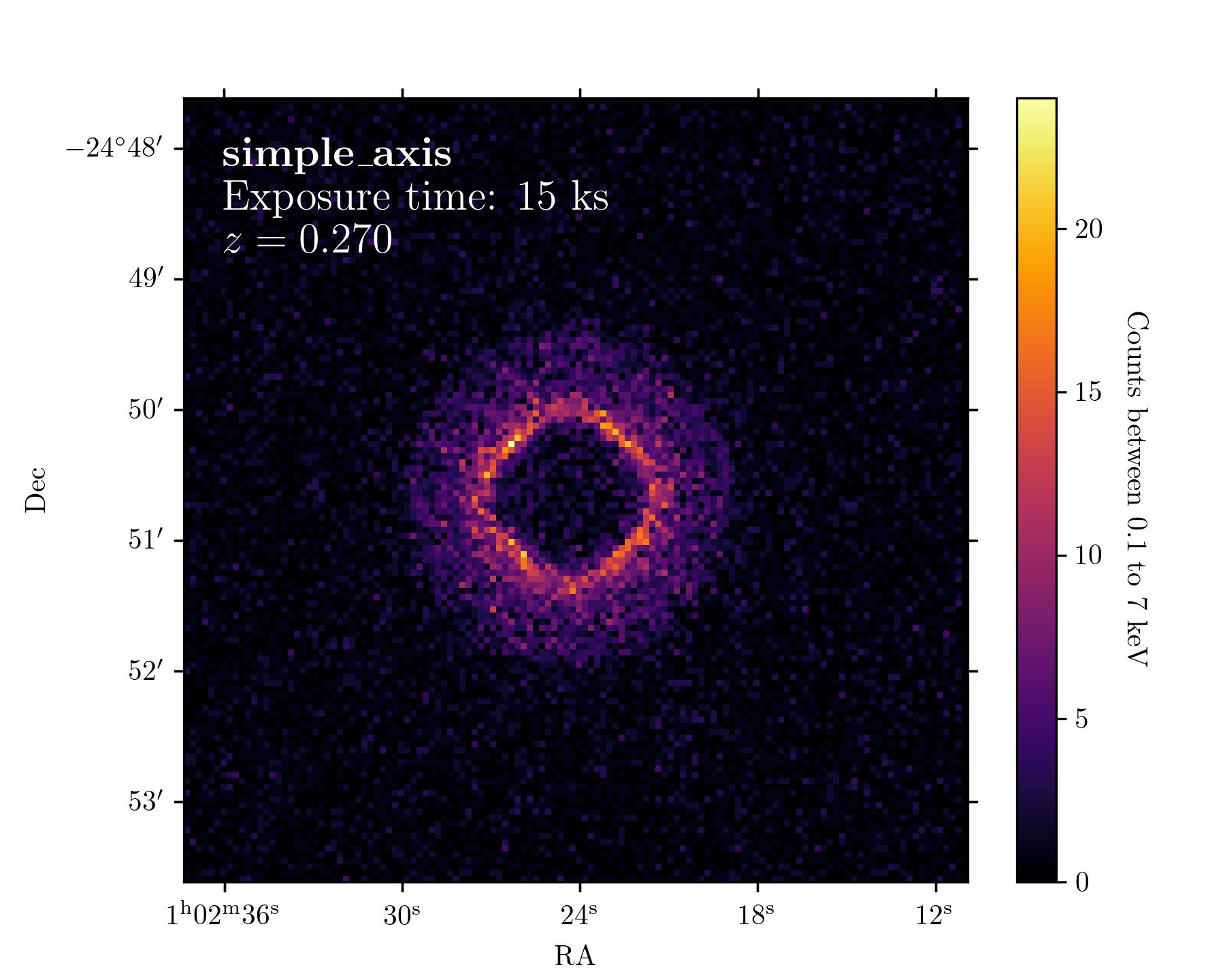}
    \includegraphics[width=\columnwidth]{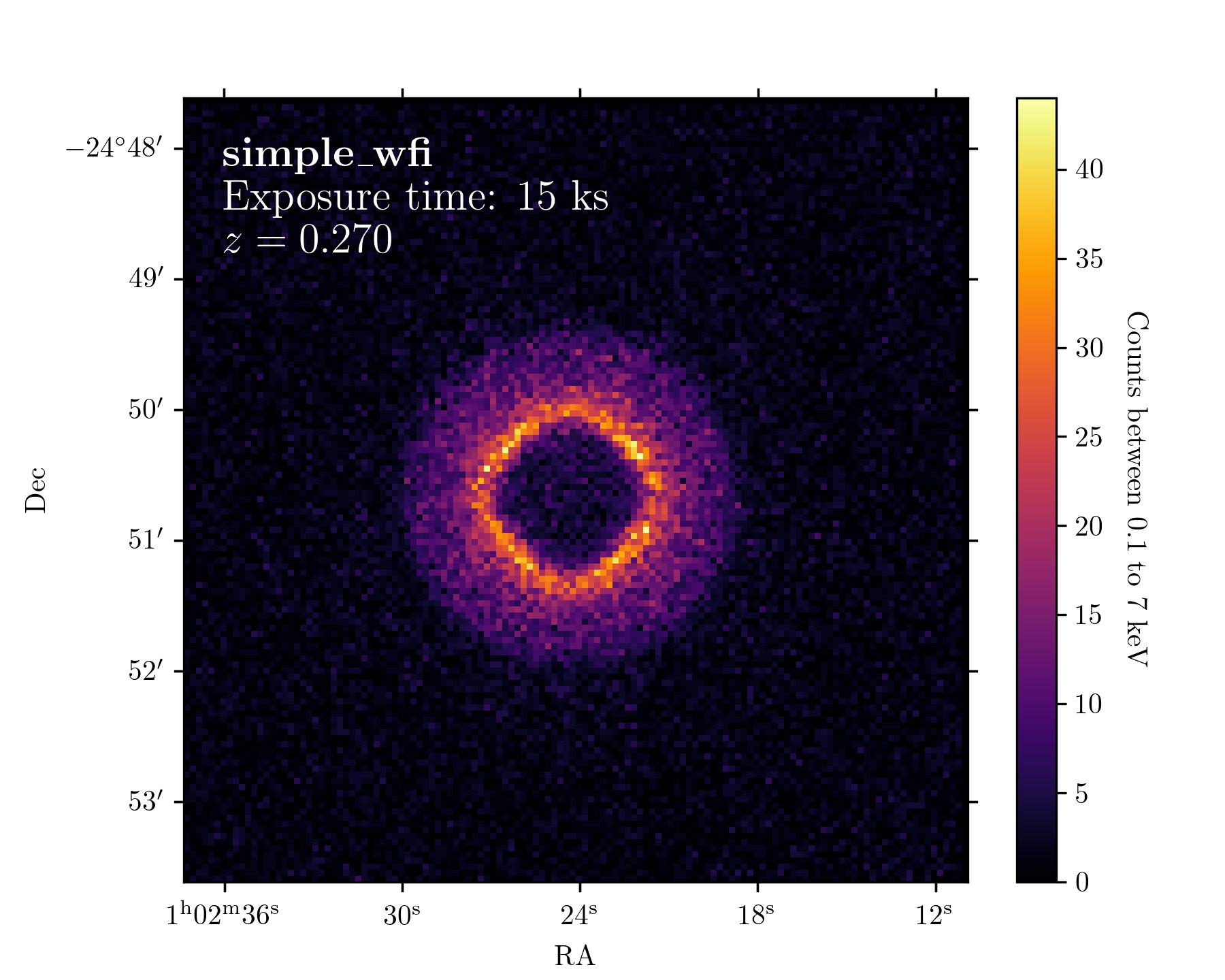}
    \caption{Mock X-ray observation of the CRpS\_M13\_P5\_D5 simulation using pyXSIM. The field of view of each image is $6'\times 6'$. For better visualization, all the images are binned to a dimension of $128 \times 128$ regardless the actual pixel size; instrumental artifacts (such as gaps between ACIS-I CCDs) are also removed.}
    \label{fig:xray_SB}
\end{figure}

As shown in Fig. \ref{fig:xray_SB}, the CRpS\_M13\_P5\_D5 case manifests a prominent thick X-ray ring. The inner edge of the ring corresponds to the high-density and turbulent bubble surface, while the outer boundary represents the shock front. We find that a long exposure time ($\sim 1$ Ms) is required for robust detection using Chandra. Conversely, AXIS and Athena could achieve a superior signal-to-noise ratio with significantly shorter exposure times ($\sim$ 15 ks).
On the other hand, mock observations of CRpS\_M12\_P5\_D5 show no detection using all three instruments up to 1 Ms of integration time. This is mainly due to the larger distance ($z=0.551$) and the lower gas density (because free-free emissivity $\epsilon_{\rm ff}\propto \rho^2T^{1/2}$). Observationally, to date there has been no detection of X-ray counterpart of ORCs in the archival data \citep{Norris2021PASAORC}. Our results, namely, the long exposure time required for M13 and non-detection for M12, are consistent with the currently available X-ray constraints. Deep X-ray observations, especially using the upcoming X-ray telescopes (e.g. AXIS and NewAthena), could be a powerful way to verify our model predictions.

\subsection{Is the energy injection reasonable?}
As mentioned in Section \ref{sec:parasearch} and \ref{sec:discussion}, the jets must be powerful enough to evacuate the inner atmosphere of the ICM/IGM to produce the observed features of ORCs. Otherwise, radiative cooling would cause the ambient gas to collapse back onto the bubbles and prevent the bubbles from growing to a sufficiently large size to match with the observed ORCs. It is therefore reasonable to ask: are the assumed jet power and duration reasonable and consistent with the general picture of galaxy evolution?

While the exact power and duration (or duty cycle) of AGN jets are still uncertain, the values we adopted, though being at the higher end, are within constraints obtained from the observed radio galaxies \citep[e.g.,][]{Hardcastle2019A&A,Shabala2020MNRAS}. On the other hand, recent theoretical advancements could provide hints for the origin of such strong and long-duration jets. \citet{Angles-Alcazar2017MNRAS} and \citet{Byrne2023MNRAS} have found that, in a series of Feedback In Realistic Environments (FIRE) simulations, the growth of supermassive black holes (SMBHs) is suppressed in the early phase of galaxy evolution by efficient stellar feedback at the galaxy center. Only after the galaxy has accumulated a sufficient stellar mass or has established a virialized circumgalactic medium (CGM), would the stellar gravitational potential and CGM pressure be sufficient to trap the gas inside the galaxy and feed the central SMBH. When that happens, the SMBH would go through a phase of rapid growth (called ``SMBH growth transitions" by these authors) and catch up to the observed $M_{\rm BH}$-$M_\star$ relation measured at low redshifts. The point of the transition is found to happen at around $M_\star\sim 10^{10.5}~M_\odot$ or $M_{\rm halo}\sim 10^{12}~M_\odot$.
Interestingly, \citet{YangTianyi2024MNRAS} also found that jet-mode feedback in the SIMBA cosmological simulations tends to happen when galaxies reach $M_\star\sim (1-5)\times 10^{10}~M_\odot$ regardless of the redshift. In comparison, the stellar masses of ORC host galaxies (ORC1c, ORC4c, and ORC5c) are found to be around $M_\star\sim 10^{11.5}$, and their SMBH masses are around $10^{8.7}~M_\odot$ \citep{Rupke2023arXiv2IOP}. Though the stellar masses predicted by the simulations and the observed ones do not match exactly, it is an interesting conjecture that the ORCs might be the remnants of powerful jets launched by the accretion during the SMBH growth transition. The observed SMBH masses are also sufficient to drive the powerful jets required in our scenario with an Eddington rate of $\sim 40\%$. 
If this is the case, the number density of ORCs should be coupled with the cosmic star formation or stellar mass evolution, which could be easily verified by the large-scale radio survey of current (e.g. LOFAR, ASKAP, MeerKAT) and future (e.g. SKA) telescopes in the near future.

\subsection{Caveats}
While our model could successfully explain many features of ORCs, several simplifications and caveats should be noted.

\begin{enumerate}
    \item A constant isotropic CR diffusion coefficient ($\mathbf{\kappa}=3\times 10^{28}$ cm$^2$ s$^{-1}$) is adopted for all the simulations presented in this study. In reality, the diffusion coefficient should be a function of CR energy, characteristic scale of magnetic field entanglement, turbulent velocity, and Alfven speed $v_A=B/\sqrt{4\pi \rho}$ \citep[see e.g.][]{Yan08,Sharma09,Ehlert18,Yang19}. Nevertheless, since many of the above properties remain largely unconstrained in the systems of interest ($M_{\rm vir}\sim 8 \times 10^{12} - 8\times 10^{13}$), we decide to retain the original value used in \citet{Yang19} and \citet{Lin2023MNRAS} to avoid additional complications. While the CR diffusion coefficient could influence the predicted thickness of the ORCs (since it determines how far the CRs could ``leak out" from the bubble interior), the predicted sizes of the ring are determined by the overall dynamics and should be insensitive to the choice of diffusion coefficient.
    \item As mentioned in Section \ref{sec:simsetup}, we intentionally reduce the level of refinement at $r>20$ kpc in order to save computational costs. While this enables us to simulate bubbles with large radii, and the overall dynamics of the bubbles converges with varied resolution, the hydrodynamic and thermal instabilities on the bubble surface are likely resolution dependent. Although this is not a major problem given the limited resolution of current observations \citep[e.g. 6" in][]{Norris2022MNRAS}, higher-resolution simulations are needed to allow direct comparison with future radio observations. 
    \item Re-acceleration of CRs by shocks or turbulence is not included in our simulations.
    However, as discussed in Section \ref{sec:discussion-AB}, CR re-acceleration could be important in determining the absolute brightness of ORCs in our simulations. 
    For instance, a recent work by \citet{Shabala2024PASA} considered a scenario where an external shock re-energizes aged electrons in the AGN bubbles, which can produce similar fluxes to the observed ORCs.
    \item In this work, a re-scaled temperature profile from the Perseus cluster is used (Section \ref{sec:cluster}). However, not all cool-core clusters share the same temperature profile, let alone non-cool-core clusters and galaxy groups. Since we generate the cluster density profile based on temperature, any modification to the assumed temperature profile could influence the radiative cooling rate of the ICM/IGM. Consequently, such alterations could potentially impact the mass flux that falls back onto the bubbles, thereby influencing the size of the bubbles.
\end{enumerate}

\section{Conclusions}\label{sec:conclusion}
The ORCs are newly discovered extragalacitc radio objects with unknown origin. In this work, we performed 3D CR-MHD simulations and demonstrated that end-on CRp dominated AGN bubbles is a plausible formation scenario for the ORCs. We find that, in order to produce the observed features (circular, large radius, and edge-brightened) of the ORCs, the AGN jets must be powerful enough to evacuate all the high density gas near the center of the halo, and hence lower-mass clusters or groups are more likely to host the ORCs. For successful cases, the high hadronic collision rates on the bubble surface could produce secondary particles that emit synchrotron radiation, forming limb-brightened radio morphology. The radio morphology remains consistent with ORC1 with viewing angles up to $\sim 30^\circ$ with respect to the jet axis, relieving the requirement of perfect alignment in the AGN scenario discussed in previous studies. We discussed about the possibility of detecting X-ray counterpart emission using deep or upcoming X-ray observations. 
We further postulate that under the AGN bubble scenario, the formation of ORCs might be connected to the SMBH growth transition found in recent cosmological simulations.

In the future, we aim to further calculate the absolute synchrotron intensity, polarization, and spectral index of the simulated radio emission, which will enable direct comparison with the results in \citet{Norris2022MNRAS}. This will provide a more comprehensive picture and a crucial test for the AGN scenario as the origin of ORCs. 

\section{Acknowledgments}
YHL and HYKY acknowledge support from National Science and Technology Council (NSTC) of Taiwan (MOST 109-2112-M-007-037-MY3; NSTC 112-2628-M-007-003-MY3). HYKY acknowledges support from Yushan Scholar Program of the Minitry of Education (MoE) of Taiwan. We thank Ellis Owen and Alvina On for providing insightful comments on radio emission generated by CRs.
This work used high-performance computing facilities operated by Center for Informatics and Computation in Astronomy (CICA) at NTHU. FLASH was developed in part by the DOE NNSA- and DOE Office of Science-supported Flash Center for Computational Science at the University of Chicago and the University of Rochester. Data analysis presented in this paper was conducted with the publicly available yt visualization software \citep{yt} and astropy \citep{Astropy2022ApJ}. We are grateful to the yt development team and community for their support. This research has made use of NASA's Astrophysics Data Systems.

\appendix

\section{Estimating synchrotron emission from secondary electrons}\label{sec:Fnu_estimate}
Following the discussion in Section \ref{sec:discussion-AB}, here we perform a simple order-of-magnitude estimate of the absolute surface brightness of the simulated ORCs.

We consider a population of secondary electrons following a power-law spectrum with spectral index $p$ between $E_{\min}$ and $E_{\max}$, which would produce a power-law synchrotron spectrum with a spectral index $\alpha=(p-1)/2$. For each grid cell in the simulations, by assuming that the energy injection rate of the secondary electrons from the hadronic processes ($P_{\rm had}$) equals to the total power of synchrotron ($L_{\rm sync}$) and IC losses ($P_{\rm IC}$), we have
\begin{equation}
    P_{\rm had} = 6(L_{\rm sync}+P_{\rm IC}) = 6L_{\rm sync}(1+u_{\rm CMB}/u_B),
\end{equation}
where $u_{\rm CMB}$ and $u_B$ are the energy density of the CMB and the magnetic field, respectively. The coefficient of 1/6 comes from the fact that only roughly 1/6 of the energy of hadronic collisions is transferred into secondary electrons, while the remaining is lost via gamma-ray photons. The total synchrotron power integrated over all the cells across the frequency range between $\nu_{\min}$ and $\nu_{\max}$ is
\begin{equation}
    L_{\rm sync} = \int_V \epsilon_{\rm sync} dV = \int^{\nu_{\max}}_{\nu_{\min}} 4\pi j_\nu d\nu,\quad
    \begin{aligned}
        \nu_{\max} &= \gamma_{\max}^2 \nu_G \\
        \nu_{\min} &= \gamma_{\min}^2 \nu_G
    \end{aligned}
\end{equation}
where $\nu_G = 2.8\ {\rm MHz}(B/{\rm Gauss})$ is the gyro frequency, and $\gamma_{\max}=E_{\max}/(m_e c^2)$ and $\gamma_{\min}=E_{\min}/(m_e c^2)$ are the Lorentz factors of the electrons at $E_{\rm max}$ and $E_{\rm min}$, respectively. The synchrotron emission coefficient $j_\nu$ should take the following form
\begin{equation}
    j_\nu = K B^{(p+1)/2}\nu^{(1-p)/2}.
\end{equation}
By re-arranging the above and performing the integral we have
\begin{equation}
    K = \frac{P_{\rm had}}{6}\frac{3-p}{8\pi}B^{-(p+1)/2}\left(\nu_{\max}^{(3-p)/2}-\nu_{\min}^{(3-p)/2}\right)^{-1}\frac{u_B}{u_B+u_{\rm CMB}}.
\end{equation}
This gives
\begin{equation}
    j_\nu = K B^{(p+1)/2}\nu^{(p-1)/2} = \frac{P_{\rm had}}{6}\frac{3-p}{8\pi}\nu^{(1-p)/2}\left(\nu_{\max}^{(3-p)/2}-\nu_{\min}^{(3-p)/2}\right)^{-1}\frac{u_B}{u_B+u_{\rm CMB}}.
\end{equation}
Finally, we calculate the flux as
\begin{equation}
    F_{\rm \nu,obs} = \frac{L_{\rm \nu, rest}}{4\pi D_{\rm L}(z)},
\end{equation}
where $\nu_{\rm rest} = \left( 1+z \right) \nu_{\rm obs}$ due to cosmological redshift, and $D_{\rm L}(z)$ is the luminosity distance assuming $H_0 = 70 {\rm ~km/(s \cdot Mpc})$ and $\Omega_{\rm m}=0.3$ with a flat $\Lambda$CDM cosmology.

\section{pyXSIM simulations}\label{sec:pyxsim_details}
We use pyXSIM \citep{ZuHoneHallman2016ascl} to make mock X-ray observations from the simulated gas distributions. We model the ICM as thermal plasma in collisional ionization equilbrium (CIE) with properties described by the APEC model \citep{APEC_Smith2001ApJ} assuming a metallicity of $Z=0.3Z_\odot$, as commonly assumed for galaxy clusters. We use the ``tbabs" model with neutral hydrogen column density $N_{\rm H}=4\times 10^{20}~{\rm cm^{-2}}$ to model the foreground Galactic absorption. M13 is placed at ${\rm (R.A.,~Dec.)}=(315.75^\circ,~ 62.00^\circ)$, $z=0.27$ and M12 is placed at ${\rm (R.A.,~ Dec.)}=(15.60^\circ, ~-24.84^\circ)$, $z=0.551$. The spectral range covers the energy range from 0.1 to 7 keV.

For the instruments, we utilize the \href{https://hea-www.cfa.harvard.edu/soxs/users_guide/instrument.html#making-simple-square-shaped-instruments}{\tt make\_simple\_instrument} function to eliminate instrumental artifacts while obtaining the key properties, such as effective area, for a specified instrument. We set the field of view (FOV) and image dimensions to be $6'\times 6'$ and $128\times 128$, respectively. For simplicity, we disabled dithering ({\tt no\_dither=True}) but keep the instrumental background ({\tt no\_bkgnd=False}). 
The instrument response files we used are:
\begin{enumerate}
    \item Chandra ACIS-I: {\tt acisi\_aimpt\_cy22.rmf, acisi\_aimpt\_cy22.arf}
    \item AXIS: {\tt axis\_ccd\_20221101.rmf, axis\_onaxis\_20221116.arf}
    \item Athena WFI: {\tt athena\_wfi\_sixte\_v20150504.rmf, athena\_sixte\_wfi\_wo\_filter\_v20190122.arf}
\end{enumerate}

\begin{figure*}
    \centering
    \includegraphics[width=0.32\textwidth]{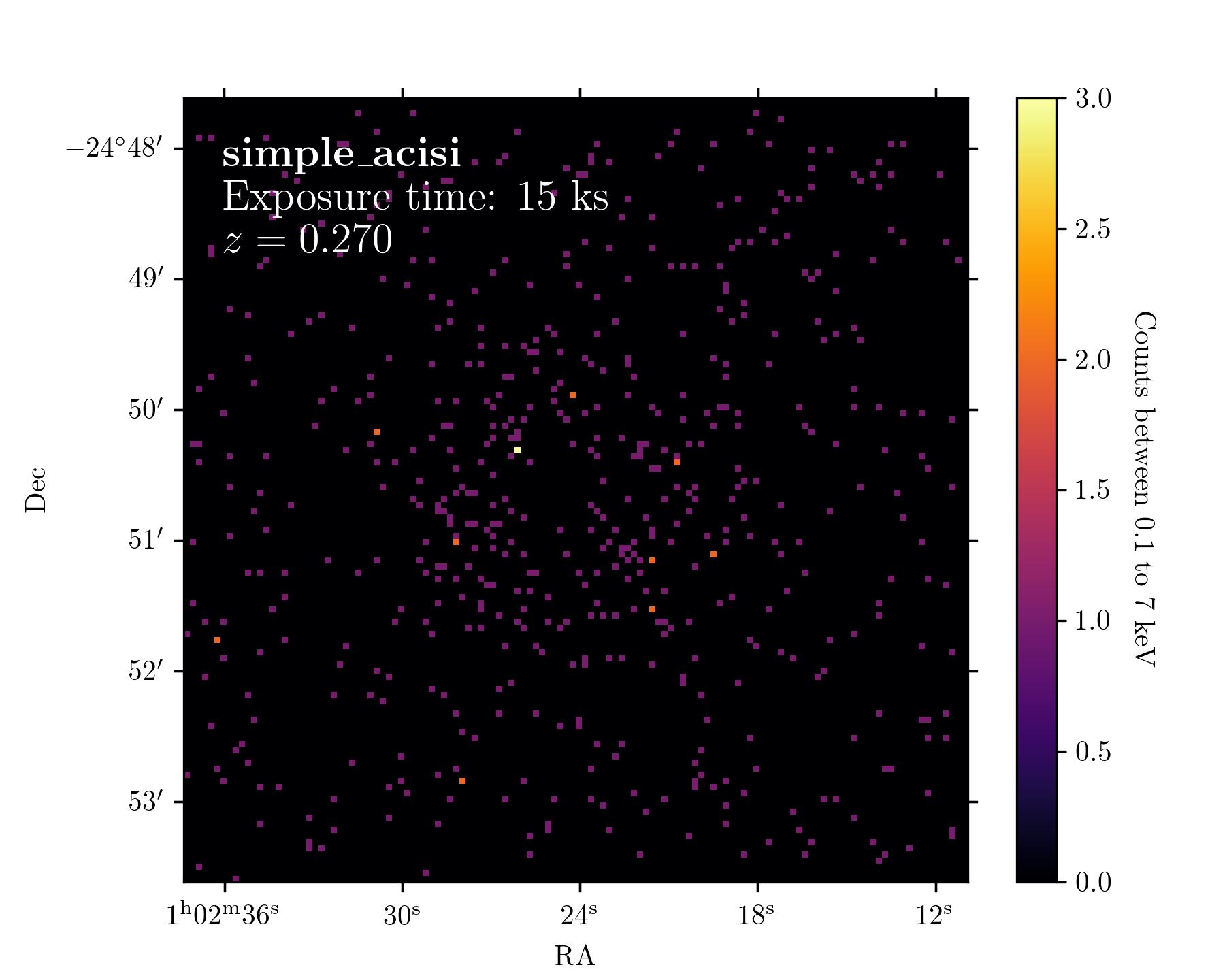}
    \includegraphics[width=0.32\textwidth]{ORCM13_simple_axis_img_0100_15ks.jpg}
    \includegraphics[width=0.32\textwidth]{ORCM13_simple_wfi_img_0100_15ks.jpg}
    \includegraphics[width=0.32\textwidth]{ORCM13_simple_acisi_img_0100_1000ks.jpg}
    \includegraphics[width=0.32\textwidth]{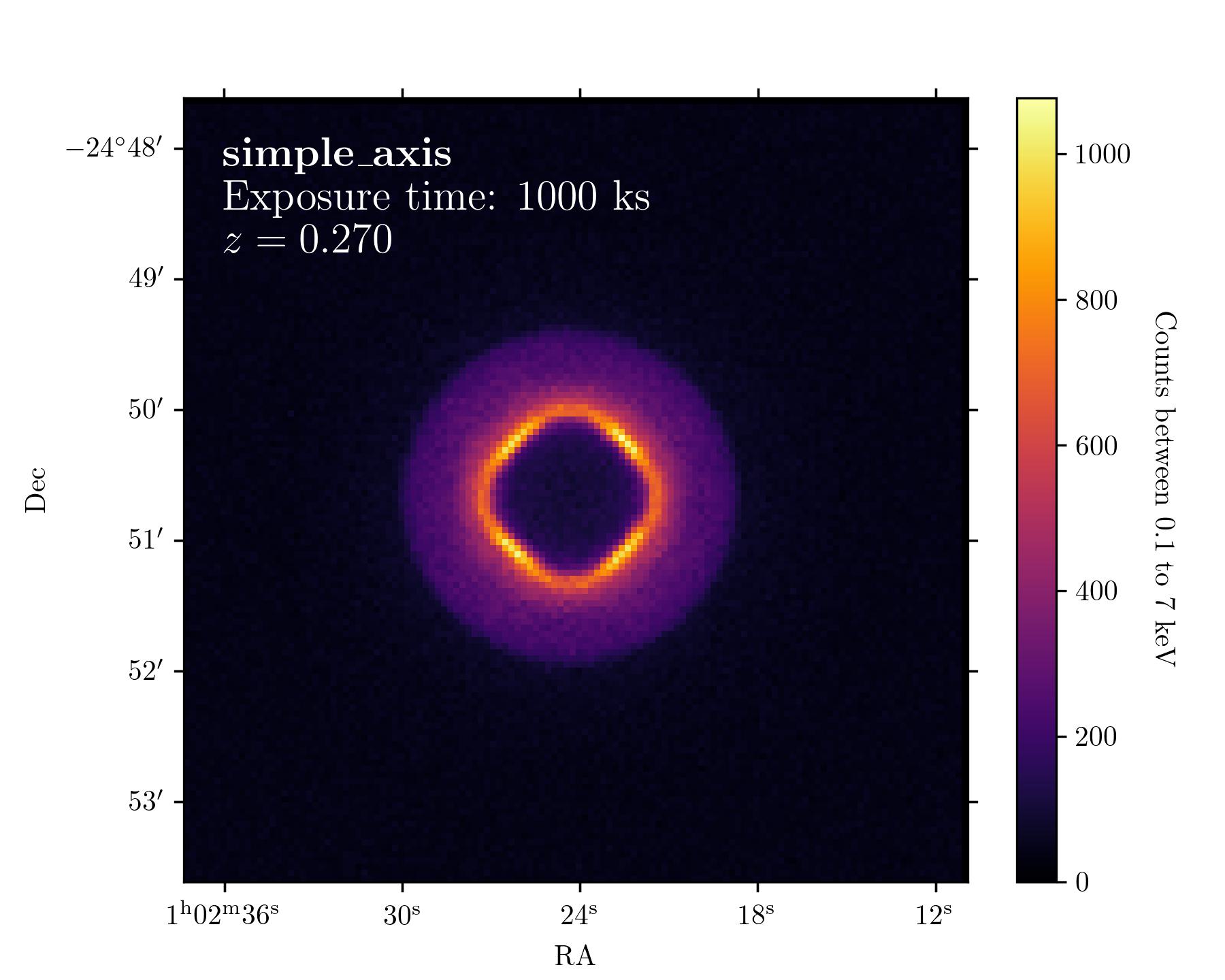}
    \includegraphics[width=0.32\textwidth]{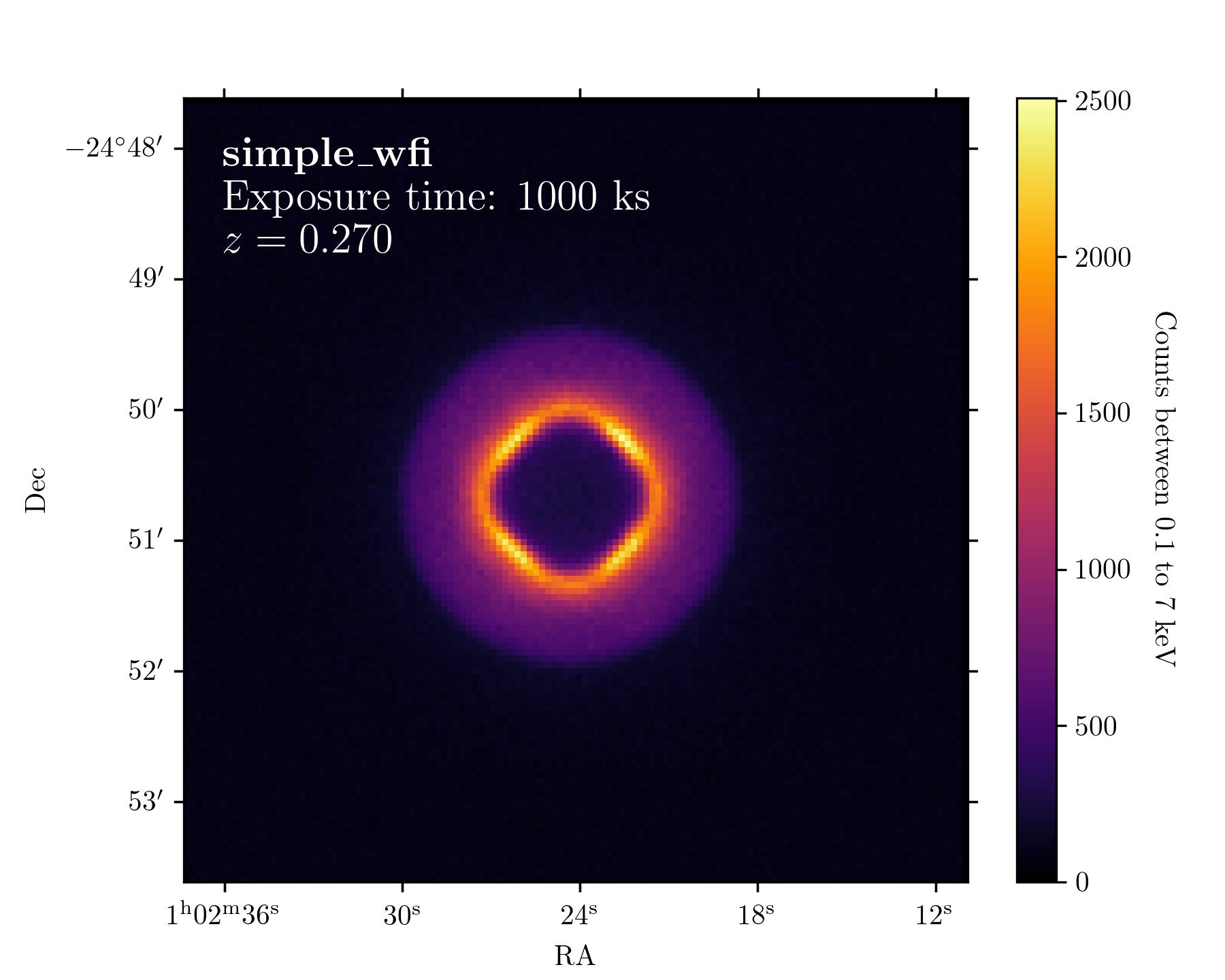}
    \caption{Mock X-ray images of the CRpS\_M13\_P5\_D5 simulation using pyXSIM, assuming exposure times of 15 ks (top row) and 1 Ms (bottom row), respectively. Images from left to right are obtained using Chandra, AXIS, and Athena response files, respectively. All images are binned into a $128 \times 128$ grid for improved visualization.}
    \label{fig:mockxray_addM13}
\end{figure*}

\begin{figure*}
    \centering
    \includegraphics[width=0.32\textwidth]{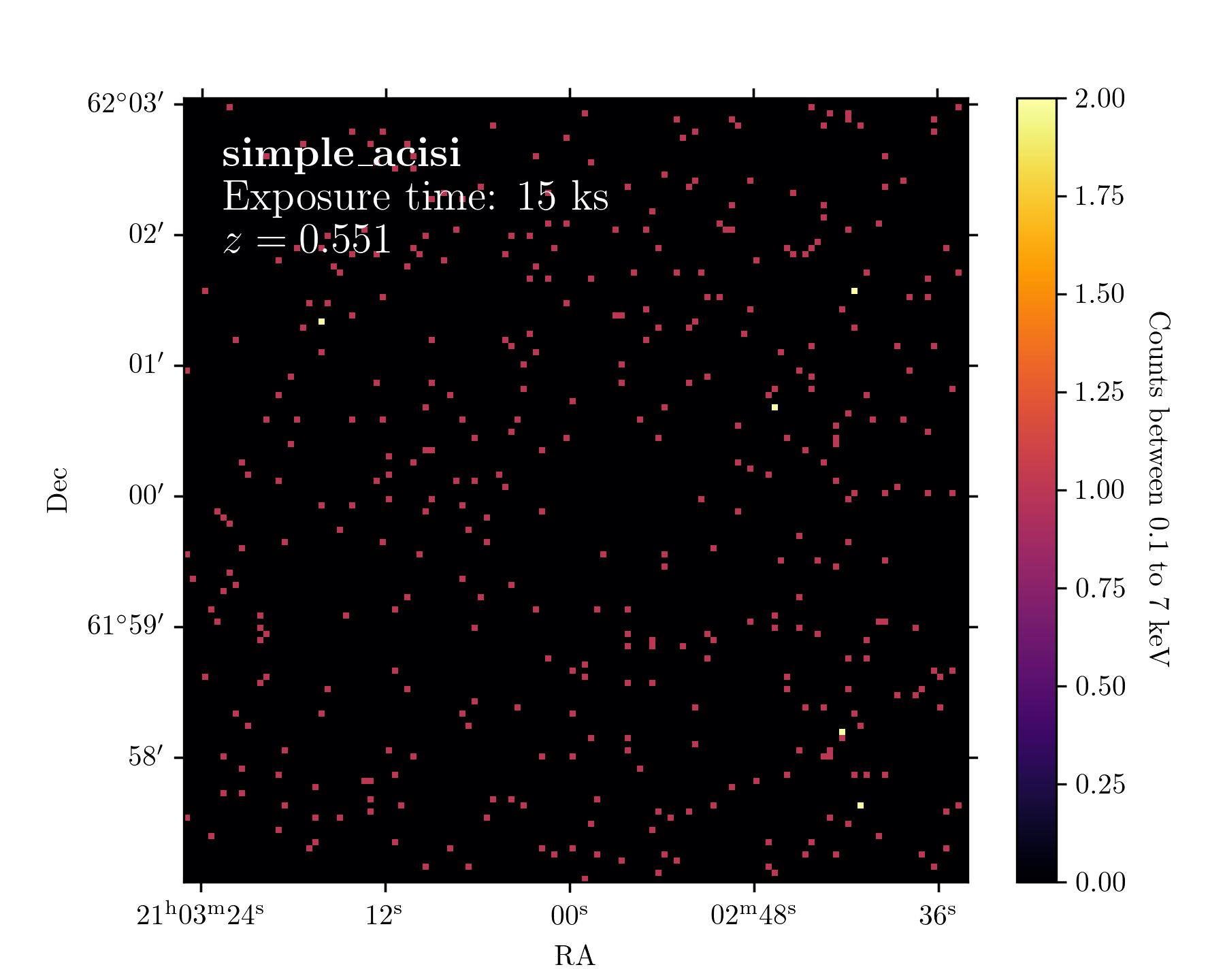}
    \includegraphics[width=0.32\textwidth]{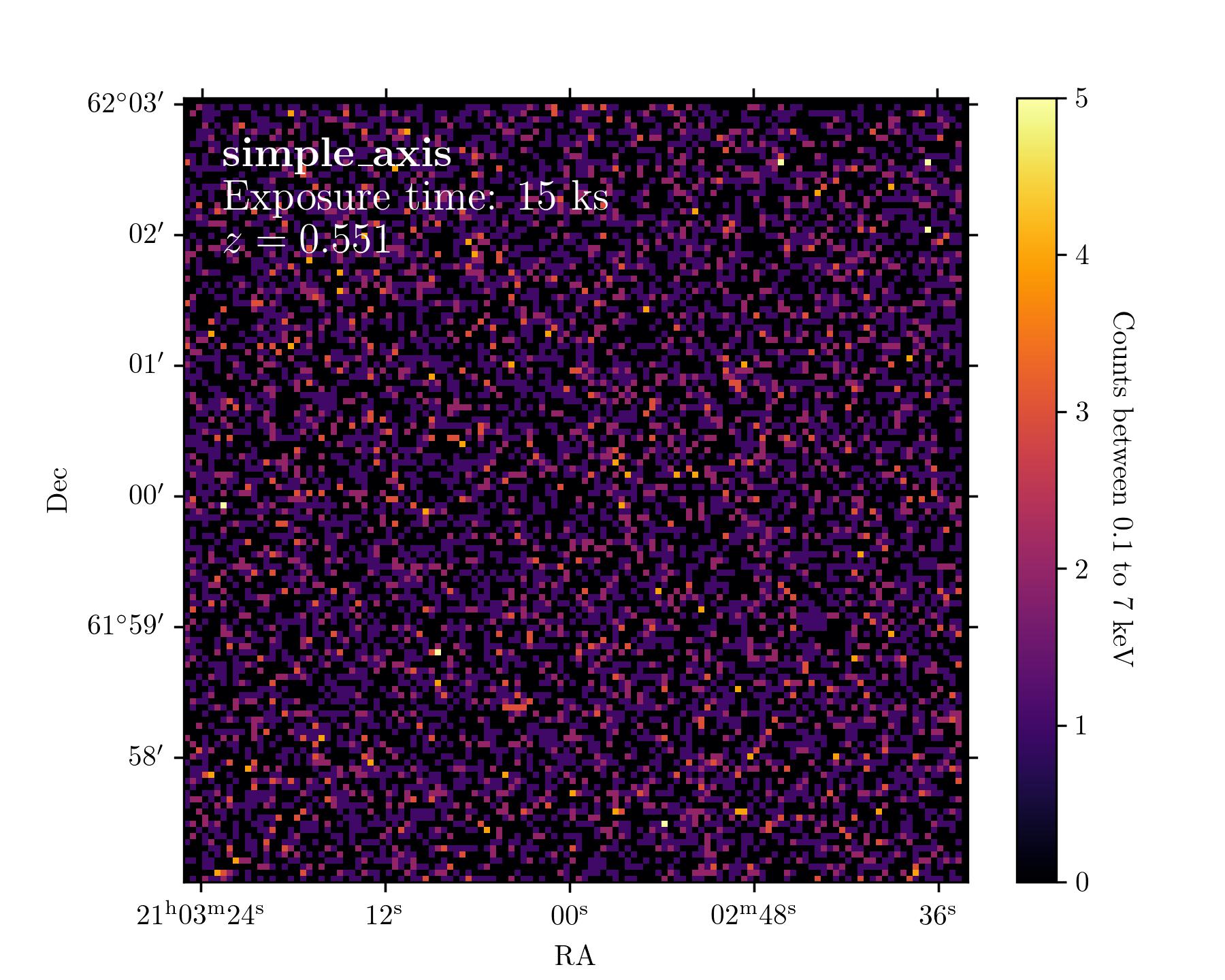}
    \includegraphics[width=0.32\textwidth]{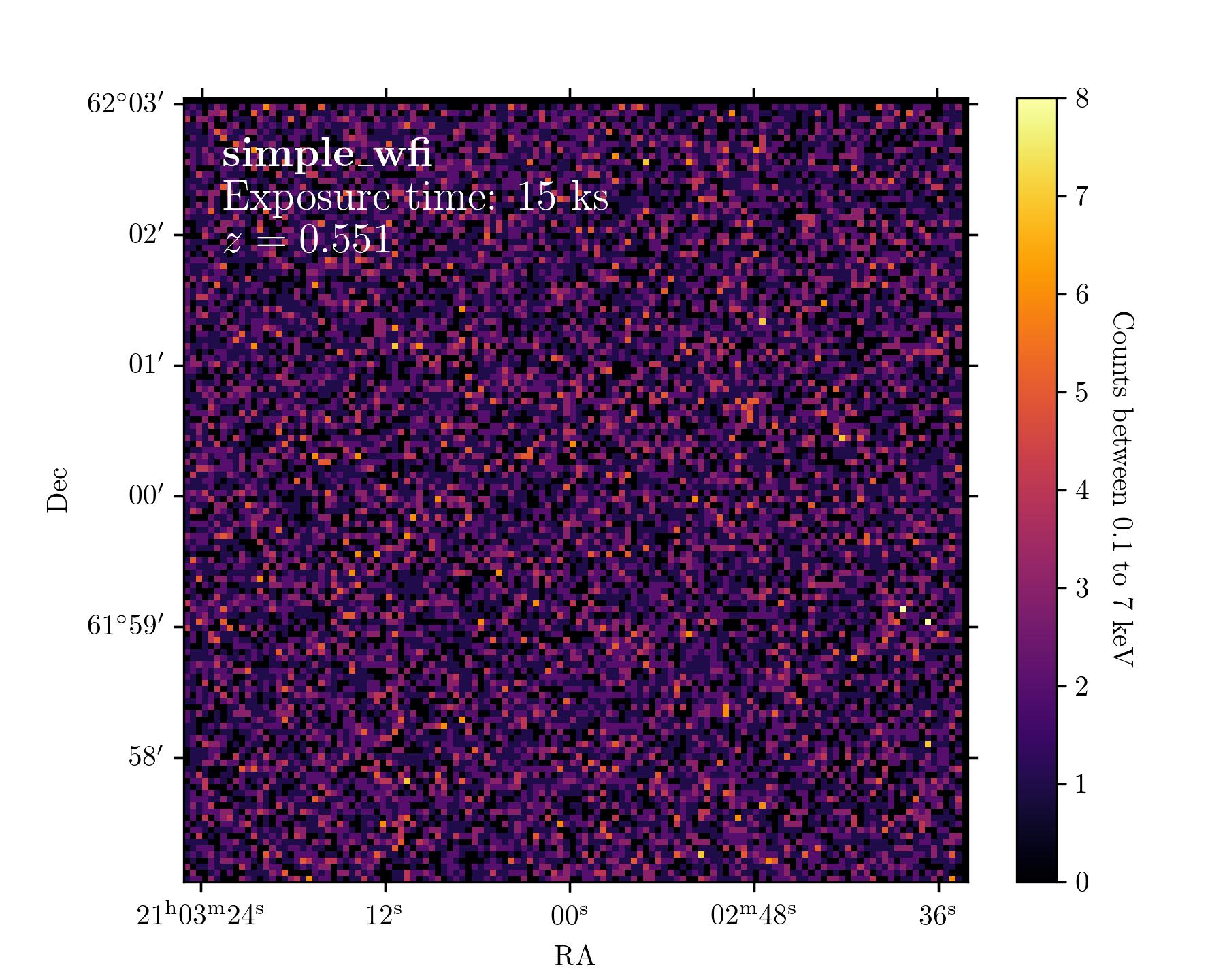}
    \includegraphics[width=0.32\textwidth]{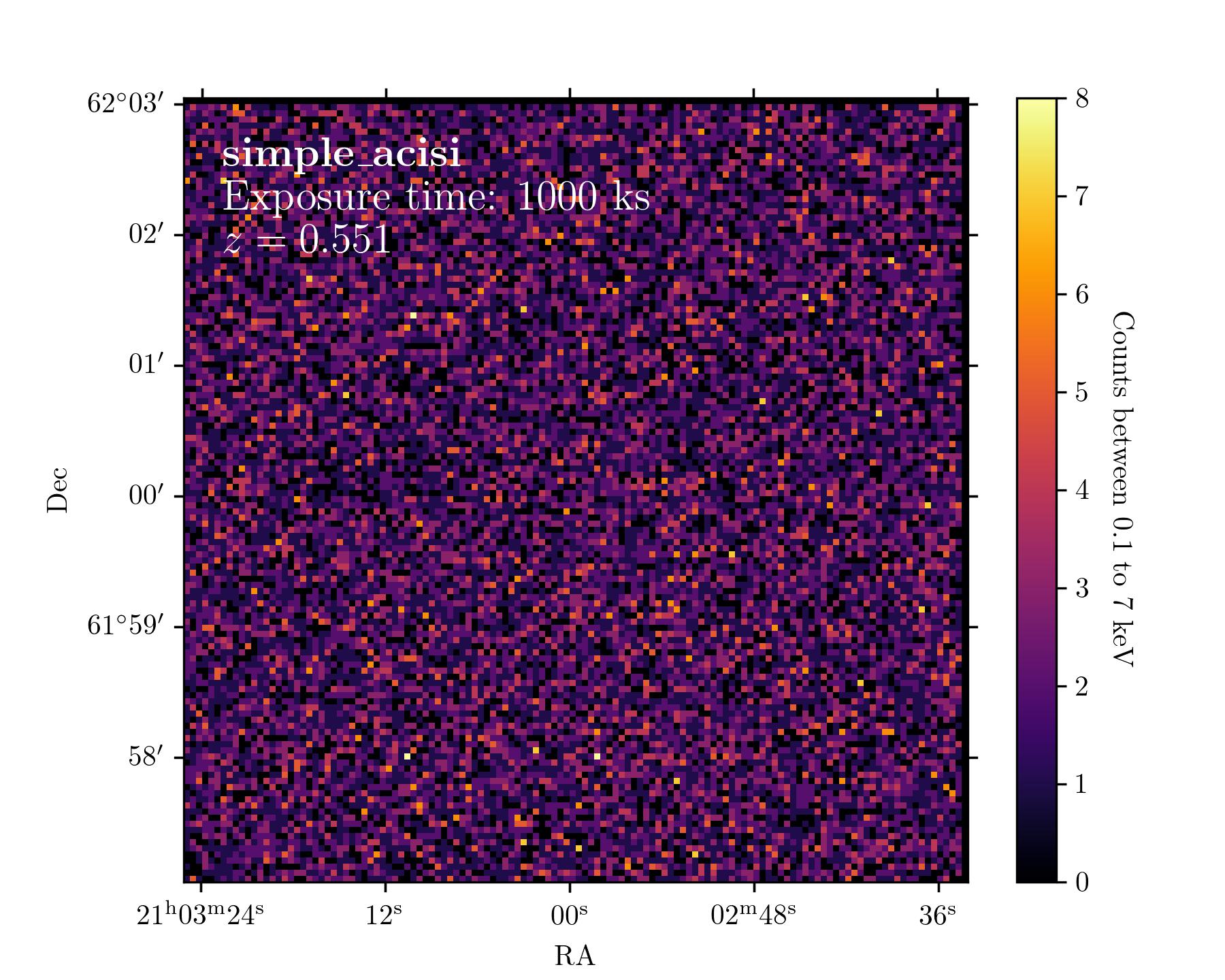}
    \includegraphics[width=0.32\textwidth]{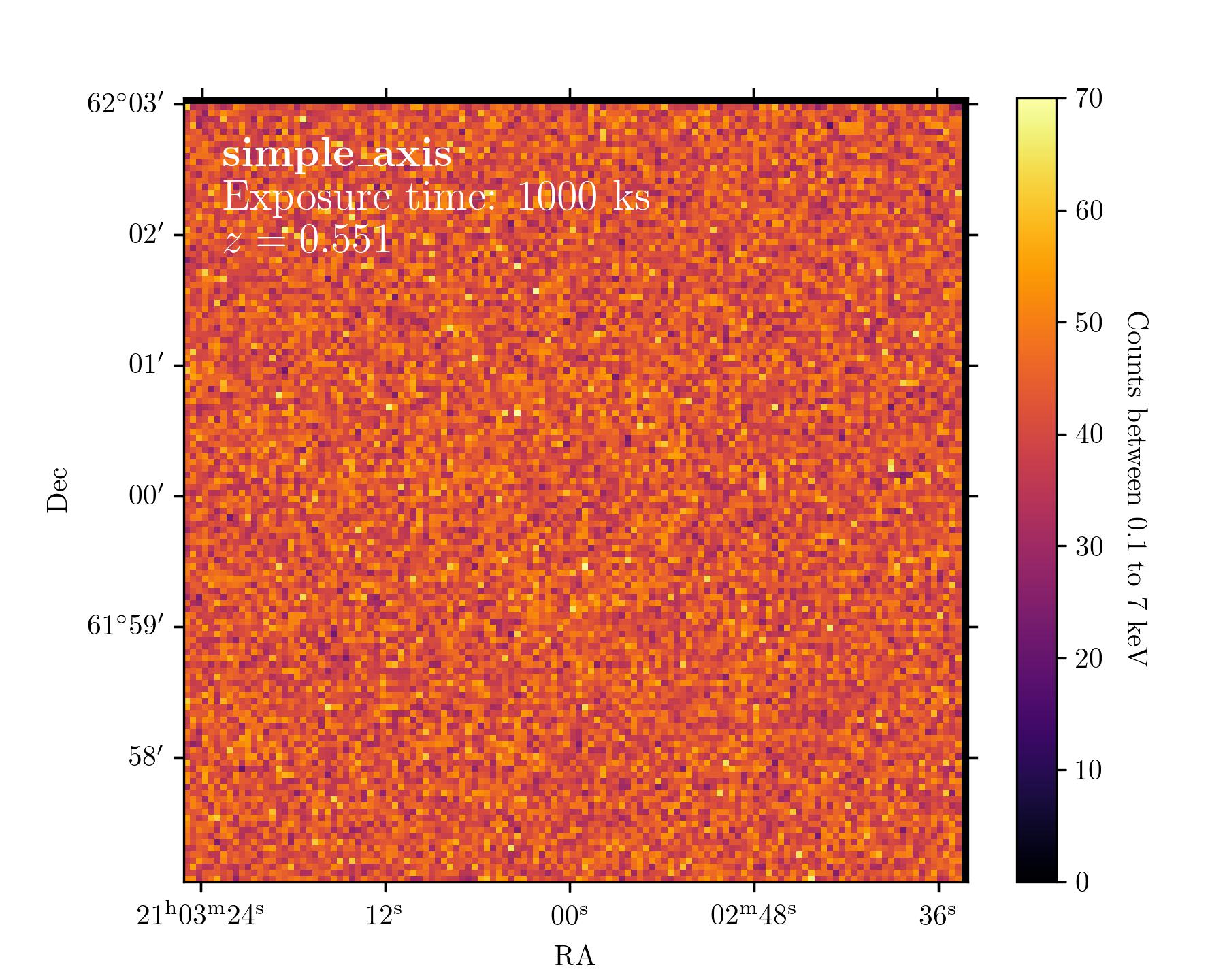}
    \includegraphics[width=0.32\textwidth]{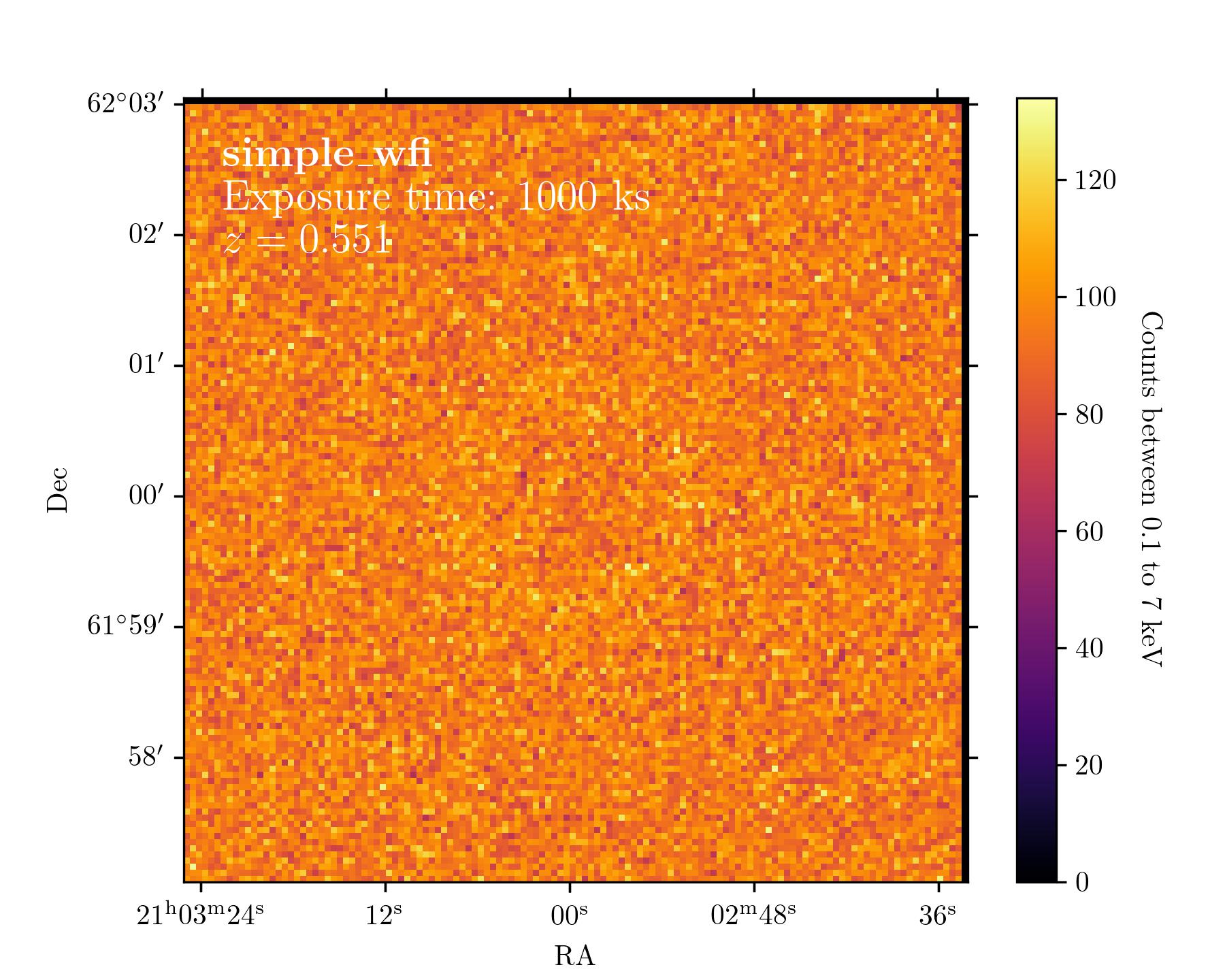}
    \caption{Same as Fig \ref{fig:mockxray_addM13}, but for the CRpS\_M12\_P5\_D5 simulation.}
    \label{fig:mockxray_addM12}
\end{figure*}

In Fig. \ref{fig:mockxray_addM13}, we present mock X-ray images obtained with ACIS-I, AXIS, and WFI, considering integration times of 15 ks and 1 Ms. As discussed in Section \ref{sec:mockXray}, detecting M13 poses a challenge for Chandra but is easily achievable for upcoming X-ray telescopes. Conversely, observing M12 (Fig. \ref{fig:mockxray_addM12}) proves to be challenging even for the next generation X-ray telescopes due to its greater distance and lower surface brightness.





\bibliography{PASPsample631}{}
\bibliographystyle{aasjournal}



\end{document}